\documentclass[submitting]{nst}
\usepackage{subfigure,dcolumn}
\usepackage[T2A,T1]{fontenc}
\usepackage[russian,english]{babel}
\usepackage{appendix}
\usepackage{placeins}

 \usepackage{listings}
\usepackage{pbox,calc}
\begin{document}

\title{Iterative Bayesian Monte Carlo for nuclear data evaluation}\thanks{This research has been supported by the Paul Scherrer Institute through the NES/GFA-ABE Cross Project.}

\author{E. Alhassan}
\email[erwin.alhassan@psi.ch]{}
\affiliation{Laboratory for Reactor Physics and Thermal-Hydraulics, Paul Scherrer Institute, 5232 Villigen, Switzerland}
\affiliation{Division Large Research Facilities (GFA), Paul Scherrer Institute, Villigen, Switzerland}
\author{D. Rochman}
\email[dimitri-alexandre.rochman@psi.ch]{}
\affiliation{Laboratory for Reactor Physics and Thermal-Hydraulics, Paul Scherrer Institute, 5232 Villigen, Switzerland}
\author{A. Vasiliev}
\affiliation{Laboratory for Reactor Physics and Thermal-Hydraulics, Paul Scherrer Institute, 5232 Villigen, Switzerland}
\author{M. Wohlmuther}
\affiliation{Division Large Research Facilities (GFA), Paul Scherrer Institute, Villigen, Switzerland}
\author{M. Hursin}
\affiliation{Laboratory for Reactor Physics and Thermal-Hydraulics, Paul Scherrer Institute, 5232 Villigen, Switzerland}
\affiliation{École Polytechnique Fédérale de Lausanne, Lausanne, Switzerland}
\author{A.J. Koning}
\affiliation{Nuclear Data Section, International Atomic Energy Commission (IAEA), Vienna, Austria}
\affiliation{Division of Applied Nuclear Physics, Department of Physics and Astronomy, Uppsala University, Uppsala, Sweden}
\author{H. Ferroukhi}
\affiliation{Laboratory for Reactor Physics and Thermal-Hydraulics, Paul Scherrer Institute, 5232 Villigen, Switzerland}


\begin{abstract}
In this work, we explore the use of an iterative Bayesian Monte Carlo (IBM) procedure for nuclear data evaluation within a Talys Evaluated Nuclear data Library (TENDL) framework. In order to identify the model and parameter combinations that reproduce selected experimental data, different physical models implemented within the TALYS code, were sampled and varied simultaneously to produce random input files with unique model combinations. All the models considered were assumed to be equal a priori. Parameters to these models were then varied simultaneously using the TALYS code system to produce a set of random ENDF files which were processed into x-y tables for comparison with selected experimental data from the EXFOR database within a Bayesian framework. To improve our fit to experimental data, we iteratively update our 'best' file - the file that maximises the likelihood function - by re-sampling model parameters around this file. The method proposed has been applied for the evaluation of p+$^{111}$Cd and $^{59}$Co between 1 - 100 MeV incident energy region. Finally, the adjusted files were compared with experimental data from the EXFOR database as well as with evaluations from the TENDL-2017 and JENDL-4.0/HE nuclear data libraries. 
\end{abstract}


\keywords{Iterative Bayesian Monte Carlo (IBM); Nuclear reaction models; model parameters; adjustments; Bayesian calibration; nuclear data; TALYS.}

\maketitle

\section{Introduction}
The use of nuclear reaction models combined with experimental data and Bayesian statistical inference has gain prominence in nuclear data evaluation (especially in the fast energy region) over the past decade or so. These techniques have been developed partly, in order to overcome the assumption of linearity used with the Generalized Least Squares (GLS) methods~\cite{bib:1a,bib:1b} used in nuclear data evaluation, and also, because of the increasing availability of computational resources which now makes large Monte Carlo calculations possible. Examples of nuclear data and covariance evaluation methods based on microscopic experimental data and statistical inference include the Total Monte Carlo (TMC) method presented in Ref.~\cite{bib:1}, the Bayesian Monte Carlo~\cite{bib:2,bib:2a}, the filtered Monte Carlo~\cite{bib:3}, the Backward-Forward Monte Carlo (BFMC)~\cite{bib:4}, the Unified Monte Carlo (UMCB-G and UMC-B)~\cite{bib:5,bib:6} and the combination of Total Monte Carlo and the Unified Monte Carlo (TMC + UMC-B) methods as presented in Ref.~\cite{bib:8}. Successful applications of the BMC and BFMC methods with respect to integral experiments have been presented in Refs.~\cite{bib:10b,bib:10c,bib:10d}. Also available is the Monte Carlo Bayesian Analysis (MOCABA) method which uses Bayesian updating algorithms for the adjustment of nuclear data based on integral benchmark experiments~\cite{bib:9a}. A similar approach based on Bayesian Monte Carlo is presented in Ref~\cite{bib:7}.

One underlying assumption of the Monte Carlo-based evaluation methods using microscopic experiments (as presented above) is that, the source of uncertainty in nuclear data is as a result of our imperfect knowledge of the parameters to nuclear reaction models~\cite{bib:4}. Therefore, by varying the input parameters to these models, one could overlap most of (or selected) experimental data. However, comparisons between model calculations and experiments most often reveal that nuclear reaction models are still deficient and therefore are unable to reproduce experimental data. In some cases, the models appear to be completely off the experimental data available i.e. they are not able to reproduce even the shape of the experimental data~\cite{bib:12,bib:9}. An example is the $^{59}$Co(p,2np) channel between 1 and 100 MeV, where large deviations were observed between model calculations and experiments. One approach (assuming the models were perfect but with uncertain parameters), has been to widen the model parameter space in order to increase the likelihood of drawing parameter combinations that can better reproduce experimental data as carried out in Refs.~\cite{bib:2,bib:7}. However, as observed in Refs.~\cite{bib:8,bib:7}, increasing the parameter space could lead to situations where a combination of model parameters are being drawn from a region of the parameter space where the likelihood is low. This normally leads to a situation where very low or insignificant file  weights are assigned to a large number of the random nuclear data files produced as observed in Refs.~\cite{bib:8, bib:7}. 

One approach has been to attribute the inability of models to reproduce experimental data, to the presence of model defects and to try to incorporate these defects in evaluations in a statistically rigorous way as presented in Refs.~\cite{bib:9,bib:10,bib:11}. In Refs.~\cite{bib:2,bib:4} for example, the likelihood functions were modified in order to take into account, the effects of these model defects. While the efforts at including the effects of model defects into the evaluation process is very commendable, we believe  that, since the model space has been left largely unexplored~\cite{bib:12,bib:13}, by exploring the model (and model parameter) space, we can be able to identify the 'optimal' model combination with its parameter set, that can better reproduce available microscopic experimental data. A more statistically rigorous approach for including the model uncertainties would be to carried out a Bayesian Model Averaging over all or a selection of the model combinations available as presented in Ref.~\cite{bib:13a}. Bayesian Model Averaging (BMA) was however, not carried out in this work. 

Also, in the evaluation of nuclear data for a general-purpose library, one often needs to consider different types of experimental data such as the reaction cross sections, residual production cross sections, angular distributions as well as double differential cross sections, among others. In most evaluations however, these experimental types are considered separately. In some cases, individual channels are evaluated separately and then combined into an evaluation. While this approach would normally result in a good fit for a particular channel or experimental type, it could lead to inconsistencies in evaluated files as the sum rules must be obeyed in evaluations. New modern nuclear data libraries such as the TENDL library~\cite{bib:13}, insists on having the same physics models (and parameters) for evaluating the reaction cross section as well as for other types such as the residual production cross sections and angular distributions. In line with the TENDL library’s philosophy of reproducibility, automation, quality assurance and completeness~\cite{bib:13}, our goal is to therefore identify a file that performs globally well with all its information stored in a single TALYS input file. This was achieved in this work by optimizing our model calculations to three experimental data types: (1) reaction cross sections, (2) residual production cross sections and (3) the elastic angular distributions. 

Additionally, in order to improve our fits (to experimental data), we update our 'best' file by re-sampling around this file in an iterative fashion, each time using the previous 'best' file (with its model and parameter set) as the new central file. We believe that after a number of iterations (within the limits of the considered models), convergence would be reached. The convergence criterion used is the relative difference of the value of the maximum likelihood estimate between the last two to five iterations. In this work however, because of computational resource constraints, the last two iterations have generally been used. Once the final 'best' file is chosen, we infer the associated uncertainties to this file by re-sampling model parameters around this file. In this work, we have applied our iterative Bayes methodology (together with the variation of models and their parameters), for the evaluation and adjustment of p+$^{59}$Co and p+$^{111}$Cd between 1-100 MeV. Proton induced reactions are important for several applications. This includes, proton therapy for the treatment of cancer, medical radioisotope production, accelerator physics, among others. Proton data are also needed in the design and analysis of sub-critical reactor systems such as the proposed MYRRHA reactor (Multi-purpose hYbrid Research Reactor for High-tech Applications), which would make use of spallation reactions in order to provide a source of external neutrons for its sub-critical core~\cite{bib:14a}.

\section{Methods}

\subsection{Model Calculations}
Model calculations were performed using the TALYS version 1.84~\cite{bib:14} code. TALYS is a state of the art nuclear reactions code used for the predictions and analysis of nuclear reactions for a number of incident particles. These particles include neutrons, protons, deuterons, tritons, $^{3}$He- and $\alpha$ particles within the 1 keV to 200 MeV energy range~\cite{bib:14}. As stated in Refs.~\cite{bib:12,bib:27}, a single nuclear reaction calculation usually involves several models connected with each other in a nuclear reaction code such as TALYS. This is because, different models are normally utilized for the computation of cross sections for the different parts of the incident particle spectrum. The main model types used in nuclear reaction codes are the optical models, the pre-equilibrium and the compound nucleus models~\cite{bib:27}. Also available are the direct and fission models. In the case of the optical model for example, there are both phenomenological and  microscopic or semi-microscopic approaches implemented in TALYS. Using any one of these optical models, in combination with other models, usually leads to different TALYS outputs. 
%
%
In Table~\ref{models_varied}, similar to Ref.~\cite{bib:13}, the models which were randomly varied within the TALYS code are listed. A total of 52 different physical models were varied in this work. As can be seen in Table~\ref{models_varied}, there are 4 pre-equilibrium models, 6 level density models and 8 gamma-strength function models, among others, available in the TALYS code. 

 \begin{table}[b]
 \centering
  \caption{Selected models varied in this work showing the number of different models per each model type as implemented in TALYS. PE denotes the pre-equilibrium model. In some cases, sub-models or components of some models have also been varied. JLM refers to the Jeukenne-Lejeune-Mahaux optical model~\cite{bib:16a}}.
  \label{models_varied}
  \begin{tabular}{lcl}  
  \toprule
   TALYS keywords  &  \pbox{20cm}{Number of \\ models} & Model Name  \\
   \midrule
preeqmode    &  4 & Pre-equilibrium (PE) \\
ldmodel      &  6 & Level density models  \\
ctmglobal    & 1 & Constant Temperature \\
massmodel    & 4 & Mass model  \\
widthmode    & 4 & Width fluctuation \\
spincutmodel & 2 &  Spin cut-off parameter  \\
gshell       & 1 & Shell effects \\
statepot     & 1 & Excited state in Optical Model \\
spherical    & 1 & Spherical Optical Model \\
radialmodel  & 2 & Radial matter densities \\
shellmodel   & 2 & Liquid drop expression \\
kvibmodel    & 2 & Vibrational enhancement \\
preeqspin    & 3 & Spin distribution (PE) \\
preeqsurface & 1 & Surface corrections (PE) \\
preeqcomplex & 1 & Kalbach model (pickup) \\
twocomponent & 1 & Component exciton model \\
pairmodel    & 2 & Pairing correction (PE) \\
expmass      & 1 & Experimental masses \\
strength     & 8  & Gamma-strength function \\
strengthM1   & 2 & M1 gamma-ray strength function \\
jlmmode      & 4 & JLM optical model \\
\bottomrule
\end{tabular}
\end{table}

These models were varied simultaneously to create a set of random TALYS input files with unique model combinations. The parameters for each model combination were then randomly varied using the T6 code package~\cite{bib:13} to produce a set of random nuclear data files in the ENDF format (referred to as the parent generation (Gen. 0) in this work). The parent generation refers to the initial random nuclear data (ND) files generated from the variation of models (and their parameters) and signifies the initial generation from where all subsequent generations are produced. A model as used here represents either a complete nuclear reaction model or a sub-model, and in some cases, components of a model or sub-model. A model set or combination represents a vector of these models or sub-models, coupled together in the TALYS code for nuclear reaction calculations. Each TALYS input file contains a set of these models (as presented in Table~\ref{models_varied}) as well as the parameters to these models. In the case where a model is explicitly given in the input file, the default model is used. 

The energy grid size for model calculations was selected taking into consideration the availability of computational resources as well as the incident energies of the available experimental data. For the parent generation (Gen. 0) for example, a smaller bin size of 20 was used for the TALYS computation. This bin size was increased to 60 for the subsequent generations in order to improve the accuracy of TALYS results but at a higher computational cost. The random ENDF files produced were then converted into x-y tables for comparison with experimental data. Since no fission channels were considered, no fission models (or parameters) were varied in this work.

\subsection{Experimental data}
\label{selected_expts}
In the evaluation process, one normally needs to carefully analyse and select experimental data since using all the experiments from the EXFOR database~\cite{bib:15} without any selection, would normally lead to the computation of very large chi squares between model calculations and experiments. This is because of the presence of discrepant and outlier experiments. In this work, outliers are treated using a binary accept/reject approach. For example (as carried out also in Ref.~\cite{bib:7}), experiments that were observed to be inconsistent with all or most of the data sets available for a particular channel and energy range, were assigned a binary value of zero. In addition, experiments that deviates from the trend of the evaluations from the major nuclear data libraries (especially the ENDF/B-VIII.0 (if available) and the latest TENDL library release), were not considered. Experiments without uncertainties reported were penalized by assigning them a binary value of zero except in cases where the considered data set (without experimental uncertainties) is the only available experiment(s) for the energy range of interest. In this case, a 10\% relative uncertainty is assumed for each data point. Also, in a situation where these experimental data (without uncertainties reported), have been considered and reported to be of reasonable quality in Ref.~\cite{bib:22}, a relative uncertainty of 10\% was assumed. In some cases, as carried out also in Ref.~\cite{bib:22}, experiments that were found to be close to the threshold and hence, usually difficult to measure, were not considered in our optimization procedure. 

It is however known that, the selection and rejection of experiments as carried out in this work, would introduce some bias into the evaluation process. Ref.~\cite{bib:2} argues that, instead of rejecting discrepant and/or outlier experiments, subjective weights which takes into account the quality of each experimental data set, could instead, be assigned to each data set. In this way, 'bad' experiments would be assigned with smaller weights and hence contribute less to the optimization. Ref.~\cite{bib:31}, proposes the use of Marginal Likelihood Optimization (MLO) for the automatic correction of the uncertainties of inconsistent experiments. Another approach (as presented in Ref.~\cite{bib:30}), is to identify Unrecognized Sources of Uncertainties (USU) in experiments and try to include them in evaluations. These approaches were however, not utilized in this work. 

The following experimental data types were considered in the optimization procedure: (1) reaction cross sections (2) residual production cross sections and (3) elastic angular distributions. As reported in Ref.~\cite{bib:13}, these experimental data types are among the most measured for proton induced reactions. Also, the goal of this work has been, to identify model and parameter combinations that reproduces globally, these experimental data types. As proof of concept, the  proposed method has been applied for the evaluation of p+$^{59}$Co between 1 - 100 MeV incident energy range. In the case of p+$^{111}$Cd, only the reaction cross sections were considered in the optimization. All the experimental data used in this work were obtained from the EXFOR database.

\subsubsection{Case of p+$^{59}$Co}
In Table~\ref{Exp_data_angle} (see in Appendix), the experimental data for the elastic angular distributions used in our optimization, are presented. The angles considered were from 1 to 180 degrees while the incident energies considered were from 5 to 40 MeV. A total of 185 experimental data points were considered. 

In the case of the reaction cross sections, the following eight reaction channels were considered: (p,non-el), (p,n), (p,3n), (p,4n), (p,2np)g, (p,2np)m, (p,$\gamma$) and (p,xn). These channels were selected because (1) experimental data were available for them within the considered energy range and (2) we desire a general purpose file which has been optimized to many reaction channels as much as possible. The selected microscopic experimental data (i.e. for the reaction cross sections) used for optimization and data adjustments are presented in Table~\ref{Exp_data_rxns} in the Appendix. A total of 169 experimental data points were considered. 

For the residual production cross sections, the following reactions were considered: $^{59}$Co(p,x)$^{46}$Sc, $^{59}$Co(p,x)$^{48}$V, $^{59}$Co(p,x)$^{52}$Mn, $^{59}$Co(p,x)$^{55}$Fe, $^{59}$Co(p,x)$^{55}$Co, $^{59}$Co(p,x)$^{56}$Co, $^{59}$Co(p,x)$^{57}$Co, $^{59}$Co(p,x)$^{58}$Co, $^{59}$Co(p,x)$^{57}$Ni. Table~\ref{Exp_data_residuals} (in the Appendix) presents the experimental data considered for the residual production cross sections. A total of 141 experimental data points were considered.

\subsubsection{The case of p+$^{111}$Cd}
For the p+$^{111}$Cd case, only reaction cross sections were considered. This is because, no experimental data were available for the residual production cross sections in EXFOR (for p+$^{111}$Cd). Also, since only one experimental data set (at 15.235 MeV) was available (in the EXFOR database) for the entire considered energy range for the elastic angular distributions, this experimental data type was not considered. We believe that one experimental data set (as in this case) was not representative enough for the entire considered energy region. The following reaction channels were considered: (p,n), (p,2n), (p,2n)g, (p,2n)m, (p,3n) and (p,4n), where $m$ and $g$ denotes meta-stable and ground state respectively. In Table~\ref{Exp_data_Cd111} (see Appendix), selected microscopic experimental data used for the adjustment of p+$^{111}$Cd between the 1 - 100 MeV range, showing the number of data points, the EXFOR ID and the name of the first author of the measurements, are presented. 

\subsection{Bayesian calibration} 
\label{Bayes_calibration}
The ultimate goal of a Bayesian calibration is to maximize the likelihood that our model output are statistically consistent with experimental data~\cite{bib:13a1}. The first step in Bayesian calibration usually involves the selection of model input parameters, followed by the quantification of the uncertainties of these parameters, and the determination of their distributions. In this work, it was assumed that, we have no prior information on the models as well as on their parameters and hence, the models and their parameters were both sampled from uniform distributions. This assumption is entirely not true since the model types used (See Table~\ref{models_varied}), were pre-selected using expert judgment and to some model sensitivity analysis. However, we assume that all models selected are equally important. The model sensitivity analysis involved holding all models constant while changing say, the different optical models available in TALYS (for example), one-at-a-time. The pre-selection of models was carried out in order to limit the model space because of computational resource limitations. Furthermore, it was observed in this study that, some model combinations give model outputs with non-smooth curves which look unphysical and therefore are excluded from subsequent model runs. It should be noted however that, this is isotope dependent and should therefore be done on a case-by-case basis. We do understand that by using expert opinion, we introduce a user bias into the initial model selection process. To exclude this bias, all the models available in the TALYS code, could have been used but at a much higher computational cost. A more detailed study on the use of Bayesian model selection and the Occam’s Razor for the selection of model combinations within a Total Monte Carlo framework is proposed and presented in Ref.~\cite{bib:29}. In the case of the input parameters to the models, all the parameters available within the TALYS code were considered. In this way, we were able to exclude selection bias from the process. It should be noted however that, not all the model parameters are sensitive to the cross sections of interest or to the elastic angular distributions. Also, the parameter uncertainties used in this work are default values used to generate random nuclear data files within the TMC method.

Now, suppose that we have a set of $J$ models, $\vec{M_j}$, where $j=1,2,...,J$, given a set of experimental data ($\vec{\sigma_E}$) with corresponding uncertainties ($\Delta \vec{\sigma_E}$). Each model combination also consists of a vector of $K$ model parameters $\vec{p_k}$, where $k=1,2,...,K$. A model refers to a vector of different models and sub-models as presented in Table~\ref{models_varied} while the parameter set represents a vector of model parameters. As previously mentioned, we assume that, all models are of equal importance a priori and that each model is characterized by a uniform prior distribution ($P(\vec{M_j},\vec{p_k})$). As carried out also in Ref.~\cite{bib:7}, we assume also that, we have no prior knowledge on the model parameters and hence, the parameters $\vec{p_k}$ were drawn from a uniform distribution. If $L(\vec{\sigma_E}|\vec{M_j},\vec{p_k})$ is our likelihood function for model $\vec{M_j}$ and parameter set $\vec{p_k}$, it can be given within the Bayesian Monte Carlo approach~\cite{bib:2} as: 
\begin{equation}
  L(\vec{\sigma_E}|\vec{M_j},\vec{p_k}) \propto exp \left(-\frac{\chi_{G,(k,j)}^2}{2} \right)
  \label{Likelihoodfun1}
\end{equation}

where $\chi_{G,(k,j)}$ is the global chi square given in Eq.~\ref{global_chi2_Avg}. Given that, $P(\vec{M_j},\vec{p_k})$ is our prior distribution and $L(\vec{\sigma_E}|\vec{M_j},\vec{p_k})$ is the likelihood function given in Eq.~\ref{Likelihoodfun1}, we can now compute our posterior distribution ($P(\vec{M_j},\vec{p_k}|\vec{\sigma_E})$) as follows:
\begin{equation}
P(\vec{M_j},\vec{p_k}|\vec{\sigma_E}) = \frac{L(\vec{\sigma_E}|\vec{M_j}],\vec{p_k}) P(\vec{M_j},\vec{p_k})}{P(\vec{\sigma_E}|\vec{M_j})} 
\label{bayes}
\end{equation}

where $P(\vec{\sigma_E}|\vec{M_j})$, which is the marginal likelihood also referred to as the model evidence, is simply a normalisation constant and therefore not considered in the optimization. Eq.~\ref{bayes} becomes:
\begin{equation}
    P(\vec{M_j},\vec{p_k}|\vec{\sigma_E}) = L(\vec{\sigma_E}|\vec{M_j}],\vec{p_k}) P(\vec{M_j},\vec{p_k})
    \label{bayes01}
\end{equation}

Based on Eq.~\ref{Likelihoodfun1}, we assign each random nuclear data file with a weight equal to the likelihood function (also called BMC weights): 
\begin{equation}
w_k=exp(-(\chi_{(G,(k,j))}^2)/2)
\label{fileweight}
\end{equation}

From Eqs.~\ref{Likelihoodfun1} and \ref{fileweight}, $\vec{\sigma_E}$ is our training (experimental) data and $\chi_{G,(k,j)}^2$, the global reduced chi square for parameter set $k$ and model combination $j$, is our single objective function obtained as a linear combination of the different reduced $\chi^2$ computed for each considered experimental data type and given by: 
\begin{equation}
    \chi_{G,(k,j)}^2 = \frac{1}{N_{\beta}} \sum_{\beta=1}^{N_{\beta}} \chi_{(k,j)}^2 (\beta)
    \label{global_chi2_Avg}
\end{equation}

where $\chi_{(k,j)}^2 (\beta)$ is the reduced chi square computed using the experimental type $\beta$ and $N_{\beta}$ is the total number of considered experimental types. For $N_{\beta}$ experimental types, Eq.~\ref{global_chi2_Avg} can be given as:
\begin{equation}
\chi_{G,(k,j)}^2 =\frac{\chi_{k}^2(xs) + \chi_{k}^2(rp) + \chi_{k}^2(DA)}{N_{\beta}}
\label{global_chi2}
\end{equation}

where $N_{\beta}$ is equal to three in this case, $\chi_{k}^2(\rm xs)$ and $\chi_{k}^2(\rm rp)$ are the reduced chi squares computed for the reaction and the residual production cross sections respectively, and $\chi_{k}^2(\rm DA)$ is the reduced chi square computed for the elastic angular distributions. Eq.~\ref{global_chi2_Avg} was computed assuming that there were no correlations between the different experimental data types considered. We note that this assumption is simplistic, since in some cases, the similar instruments or methods are used in the measurements which introduces cross correlations between these experimental types. However, these correlations are not readily available and therefore not used in this work. 

Similarly, in the computation of the individual reduced chi squares in Eq.~\ref{global_chi2} such as the $\chi_{k}^2(xs)$, the experimental data were assumed to be uncorrelated. To include experimental correlations, the generalized chi square as presented in Ref.~\cite{bib:2} should have been used. However, experimental correlations are usually scarce, incomplete and most often, not readily available. Therefore, the reduced chi square with respect to the reaction cross sections ($\chi_{k}^2(\rm xs)$) for example, was computed using Eq.~\ref{reducedchi2_xs} (similar expressions  were presented also in Refs.~\cite{bib:2} and \cite{bib:12}): 
\begin{equation}
    \chi_{k}^2(xs) = \frac{1}{N_c} \sum_{c=1}^{N_c} \frac{1}{N_m} \sum_{m=1}^{N_m}\frac{1}{N_{pt}} \sum_{i=1}^{N_pt} \bigg(  \frac{\sigma^{ci}_{T(k)} - \sigma^{cmi}_E}{\Delta \sigma^{cmi}_E} \bigg)^2
    \label{reducedchi2_xs}
\end{equation} 

Where $N_c$ is the total number of considered channels $c$, $N_m$ is the total number of experimental data sets $m$, and $N_{pt}$ is the total number of considered data points for each experimental data set; $\sigma_E^{cmi}$ and $\sigma_{T(k)}^{ci}$  are vectors of the experimental and TALYS calculated cross sections at the energy $i$, for the data set $m$, and channel $c$ respectively. Similarly, $\Delta\sigma_E^{cmi}$ is the corresponding experimental uncertainty at energy $i$, data set $m$ and channel $c$. In cases where there are no matches in energy (or in angle in the case of the elastic angular distributions) between the TALYS calculations and the considered experiments, similar to what was carried out in Refs.~\cite{bib:7,bib:12}, we linearly interpolate to fill in the missing TALYS values. The same approach as presented in Eq.~\ref{reducedchi2_xs} was applied for the computation of $\chi_{k}^2(\rm rp)$ and $\chi_{k}^2(\rm DA)$. However, in the case of the $\chi_{k}^2(\rm DA)$, we match TALYS calculations with that of experiments in both energy and angle, however, we only interpolate on the angle. 
%

In Refs.~\cite{bib:7,bib:12}, the reduced chi square presented in Eq.~\ref{reducedchi2_xs}, was computed by averaging the chi square values for each channel over all the considered experimental data points. This approach, as stated also in Ref.~\cite{bib:2}, assigns equal weights (aside their uncertainties) to all the experimental data points which may lead to a situation where an experimental data set with a large number of measurements completely dominants the goodness of fit estimations. Also, channels with many different experimental data sets but fewer measurements would be assigned smaller weights compared to those with fewer experimental data sets but with many measurements. Experimental data sets as used here refers to one or more measurements carried out for a specific energy or energy range, channel and isotope, with a unique EXFOR ID in the EXFOR database (See Table~\ref{Exp_data_rxns}). In an attempt to assign channels with many different experimental data sets with larger weights (aside their uncertainties), inline with what was carried out in Ref.~\cite{bib:2}, we averaged the chi square over each experimental data set by dividing by $N_{pt}$ as carried out in Eq.~\ref{reducedchi2_xs}. In this way, each experimental data set contributes equally to the goodness of fit estimation. Further, by averaging over each experimental data set, we in a way, combine the information from each experimental set into a single goodness of fit estimate since the measurements from each data set are known to be highly correlated. The correlations come about as a result of the fact that, usually, the same equipment as well as methods (and authors) are used or involved in these measurements. It is also known that, the addition of correlated experiments is not as effective in reducing the uncertainty in our calibration as the uncorrelated or independent experiments~\cite{bib:29b}. 

Statistical information in the posterior distribution in Bayesian estimations as presented in Eqs.~\ref{bayes} and \ref{bayes01}, can normally be summarized by computing central tendency statistics~\cite{bib:16}, where the posterior mean is used as the best estimate (with its corresponding variance). However, given a large sample size and assuming that our prior was sampled from a uniform distribution, as stated in Ref.~\cite{bib:17}, the posterior Probability density function (PDF) can be asymptotically approximated by a Gaussian PDF centered on the Maximum a Posteriori (MAP) estimate. Therefore, with reference to this work, as used also in Bayesian Model Selection (BMS)~\cite{bib:21}, the best or winning model becomes the model (and parameter) set that maximises the posterior probability which is also known as the Maximum a Posteriori (MAP) estimate (and that equals the mode of the posterior distribution), given as:
\begin{equation}
L_{MAP} =\operatorname*{arg\,max}_{m} [L(\vec{\sigma_E}|\vec{M_j},\vec{p_k})P(\vec{M_j},\vec{p_k})]
\label{MAP}
\end{equation}

where $L_{MAP}$ is our Maximum a Posteriori (MAP) estimate and the index $m$ denotes the considered models and their parameters. However, in Bayesian statistics, as stated in Ref.~\cite{bib:16}, the maximum likelihood estimation (MLE) \textit{"is a special case of Bayesian Estimation (BE) in which (1) the estimate is based on the mode of the posterior distribution, and (2) all the parameters values are equally likely (i.e there is no priors)"}. The MLE is therefore viewed as a special case of the MAP estimate in the case where a uniform distribution is assumed for the prior distribution of the parameters~\cite{bib:16}. Since in this work, we assume a uniform distribution for the models as well as for their parameters, the MLE was used. Therefore, given a uniform distribution of models and their parameters, Eq.~\ref{MAP} reduces to the maximum likelihood estimate denoted by $L_{MLE}$ and given as:
\begin{equation}
L_{MLE} =\operatorname*{arg\,max}_m [L(\vec{\sigma_E}|\vec{M_j},\vec{p_k})]
\label{MLE}
\end{equation}

From Eq.~\ref{MLE}, the random nuclear data file (with its model combination and parameter set) which maximizes the likelihood function, was considered as the file that makes the experimental data most probable for each generation. It should be noted also that, the maximum likelihood is equivalent to minimizing the chi square. The final file chosen becomes our new 'evaluation'. As stated in Ref.~\cite{bib:7}, in order to infer parameter uncertainties to this file, one would normally need to re-sample model parameters around this file using updated parameter uncertainties. In this way, the new distribution would be centered around our 'best' file. 
%

\subsection{Iterative Bayes Procedure}

In order to improve on our evaluations, we propose the use of an iterative Bayesian Monte Carlo procedure where, the 'best' file or 'evaluation' from each generation is used as the central file around which model parameters are varied to produce the next generation. 

The algorithm for the Iterative Bayesian procedure proposed in this work is presented in Table~\ref{IBM_algorithm}. The idea of the iterative procedure is to minimize the bias between our experimental observables and the corresponding model outputs in an iterative fashion. As can be seen from the Table, we start with the selection of model combinations and their parameters (including parameter uncertainties and distributions) - see Table~\ref{models_varied} for more information on the considered models. The idea was to update the parameter uncertainties for each iteration, however, since it was not possible to explore the entire parameter space (because of the computational expense involved), we think that using reduced (updated) parameter uncertainties will decrease the likelihood of converging to the optimal parameter set. Hence, the default parameter uncertainties were maintained for each iteration. According to Refs.~\cite{bib:13,bib:14}, the default parameter uncertainties were obtained by comparing scattered TALYS curves with experimental data. Also, in line with using non-informative priors as presented in Ref.~\cite{bib:7}, all parameters for each generation were drawn from uniform distributions. More information on the model parameters can be found in Ref.~\cite{bib:33}. Next, we select the energy grid and bins for the TALYS calculations. This was done taking both the target accuracy of our calculations and the computational resources available into account. Further, the energy grid was chosen such that, there were a large number of matches (in incident energy) between TALYS calculations and the corresponding experiments. 

Next, a set of random TALYS input files with different model combinations (and parameters) were generated. These input files were fed to the TALYS code within the T6 code package~\cite{bib:13} to produce a set of random nuclear data files in the ENDF format for each model and parameter combination. These files constitute the parent generation (Gen. 0). The parent generation, as stated earlier, is defined as the initial population of random nuclear data files generated from the variation of both models and their parameters. These files were processed into x-y tables for comparison with experimental data from the EXFOR database. The selection of experimental data for the considered channels has been presented earlier in section~\ref{selected_expts}. 

\begin{table}[!th]
 \centering
  \caption{Iterative Bayesian Monte Carlo (IBM) algorithm. $L(\vec{\sigma_E}|\vec{M_j},\vec{p_k})$ is the likelihood function for each iteration (or generation) $\kappa$. $\vec{\sigma_E}$ is our experimental data while $\vec{M_j}$ is our model vector and $\vec{p_k}$ is a vector of model parameters for random file $k$.}
  \label{IBM_algorithm}
  \begin{tabular}{l}  
  \toprule
   IBM algorithm    \\
   \midrule
  1: Select model combinations + parameter set (including \\ 
     \hspace{3mm} determining parameter uncertainties and their distributions) \\
  2: Select energy grid and bins for TALYS calculations \\
  3: Generate a large set of random TALYS input files by drawing  \\
  \hspace{4mm} model combinations and parameters from uniform distributions \\
  4: Execute the T6 code to produce a set of random ENDF files \\
  5: Process random ENDF files into x-y tables \\
  6: Select experimental data for considered channels \\
  7: for $\kappa$ = 1, 2, ..., do  \\
  8:  \hspace{3mm} Compute the likelihood function for each random ND file \\
  9: \hspace{3mm} Select file with 'best' model combination (BF): \\
     \hspace{7mm} \(\displaystyle  BF = \operatorname*{arg\,max}_m [L(\vec{\sigma_E}|\vec{M_j},\vec{p_k})]\) \\
  10: \hspace{3mm} Update model parameters (and their uncertainties) \\
  11: \hspace{3mm} Repeat steps 3 to 10 (until convergence) however varying \\ \hspace{8mm} only model parameters around the model set selected. \\
  12: end for \\
  13: Return a solution for Bayesian calibration \\
  14: Vary model parameters around final 'best' file to infer the \\ \hspace{5mm} uncertainties associated with this file. \\
\bottomrule
\end{tabular}
\end{table}

Next, as stated earlier, we select the 'best' model combination (and parameters) first, from the parent generation (gen. 0) by identifying the nuclear data file with the largest likelihood function value. It should be noted here that, because of computational resource constraint, the entire model space could not be covered using the brute-force approach as used in this work. The use of more efficient sampling methods is proposed for future work. 

Using the best file as our new central file or 'best' estimate, we re-sample around this file, however this time, we vary only model parameters of the models selected, to produce another set of random nuclear data files referred to as the 1st generation (Gen. 1). We then update our model parameters and repeat our model parameter variation step (steps 3 to 10 in Table~\ref{IBM_algorithm}) in an iterative fashion until we reach convergence. Note that, as previously mentioned, the parameter uncertainties used for the parent generation was used for subsequent generations. Convergence is determined by monitoring the evolution of the maximum likelihood estimate for each iteration and convergence is reached when the relative change in the maximum likelihood estimates for the last 2 to 5 iterations is within 10-20\%. Ideally, a smaller relative deviation is required but this would normally involve large amounts of computational resources or time. Therefore we tried to balance the gain in accuracy with the computational expense and time involved. The idea is to start with a large parameter space and then narrow this space as we improve on our evaluations. The motivation for using the maximum likelihood estimate as our best estimate has been presented earlier in section~\ref{Bayes_calibration}.

A point worthy of note is that, within the limits of our models in a multi-objective optimization (as in our case), a point would be reached (in the iteration process) where further variation of parameters cannot improve the fits for example in the case of the reaction cross sections, without necessarily making the other experimental data types such as the residual production cross section or angular distributions, worse-off. This condition is referred to as the \textit{Pareto optimality}~\cite{bib:26}. This is because, multi-objective optimization problems usually involves the simultaneous optimization of multiple competing objectives, which leads to trade-off solutions largely known as \textit{Pareto-optimal} solutions. A possible solution is to assign subjective weights to each objective (or experimental data type), depending on the needs of the evaluation. For example, if the target is the production of radio-isotopes, larger weights could be assigned to the residual production cross sections. This was however not carried out in this work since our goal is to provide a general purpose evaluation which compares reasonably well with all the considered experimental data types. As a rule of the thumb, when no further improvements are possible, the iteration should be stopped since the \textit{Pareto optimum} might have been reached (within the limit of the models used). Visual inspection of calculated cross sections against experimental data is sometimes needed to identify unphysical models since the chi square could miss these models if their results pass through the experimental points. 

The final 'best' file selected is then validated against available microscopic experimental data in the EXFOR database (also called differential validation) and compared with evaluations from other nuclear data libraries (if available). As mentioned previously, in order to infer uncertainties to the final 'best' file, we re-sample model parameters around this file to produce a large set of entirely random nuclear data libraries. The covariance information (which is associated to the central or 'best' file), is then contained in the distributions of these random nuclear data files. Alternatively, this information can be stored in MF31-40 (ENDF terminology). Where available, the 'best' file should also be tested against a large set of integral benchmark experiments. These benchmarks are however, not readily available in the case of proton induced reactions.


\section{Results}
\subsection{Application to p+$^{59}$Co}
The models selected (from the parent generation (Gen. 0)) are compared with the corresponding default TALYS models in the case of p+$^{59}$Co and presented in Table~\ref{models_selected}. It should be noted that, the evaluations in the proton sub-library of the TENDL-2017 library were produced using default TALYS models and parameters~\cite{bib:13}. Selected models which were found to be the same as the default models, are not presented in Table~\ref{models_selected}. These models include for example, the $widthmode$ (TALYS keyword used to invoke the models for width fluctuation corrections in compound nucleus calculations), $statepot$ (flag for using a different optical model parameterisation for each excited state in a Distorted Wave Born Approximation (DWBA) or coupled-channels calculation), $gshell$ (flag to include the damping of shell effects with excitation energy in single-particle level densities), $preeqsurface$ (flag to use surface corrections in the exciton model).

For the vibrational enhancement of the level density ($K_{vib}$) as presented in Table~\ref{models_selected}, $\delta S$ and $\delta U$ denote the changes in the entropy ($S$) and excitation energy ($U$) respectively, $t$ is the thermodynamic temperature and $A$ denotes the mass number~\cite{bib:14,bib:18,bib:19}. In the case of the spin cut-off parameter ($\sigma^2$), $a$ denotes the energy-dependent level density parameter, \~{a} is the asymptotic level density value obtained when all shell effects are damped and $c$ is the rigid body moment of inertia~\cite{bib:14,bib:18}. Gogny D1M HFB as seen from the table, represents the Hartree-Fock-Bogolyubov model with the Gogny D1M nucleon force while QRPA is the Quasi-particle Random Phase Approximation model~\cite{bib:24}. 
%

\begin{table*}
 \centering
  \caption{Selected model combination based on the parent generation (Gen. 0) in the case of p+$^{59}$Co. These models were used as the new 'best' file around which model parameters were varied to obtain the 1st generation of random nuclear data. The other models (as shown in Table~\ref{models_varied}) but not presented here were found to be the same as the default TALYS models. For the vibrational enhancement of the level density ($K_{vib}$) as presented, $\delta S$ and $\delta U$ are the changes in the entropy ($S$) and excitation energy ($U$), respectively, $t$ is the thermodynamic temperature and $A$ denotes the mass number~\cite{bib:14,bib:18,bib:19}. In the case of the spin cut-off parameter ($\sigma^2$), $U$ is the excitation energy, $a$ is energy-dependent level density parameter, \~{a} is the asymptotic level density value obtained when all shell effects are damped and $c$ is the rigid body moment of  inertia~\cite{bib:14,bib:18}. Gogny D1M HFB represents the Hartree-Fock-Bogolyubov model with the Gogny D1M nucleon force while QRPA is the Quasi-particle Random Phase Approximation model~\cite{bib:24}. $y$ and $n$ represents yes and no.}
  \label{models_selected}
  \begin{tabular}{lll}  
  \toprule
   Model name  & Selected models & default models \\
   \midrule
Pre-equilibrium & \pbox{20cm}{preeqmode 3: Exciton model - Numerical \\ transition rates with optical model \\ for collision probability}   &  \pbox{20cm}{preeqmode 2: Exciton model: Numerical \\ transition rates with energy-dependent \\ matrix element}  \\
Level density & ldmodel 2: Back-shifted Fermi gas model &  \pbox{20cm}{ldmodel 1: Constant temperature \\ + Fermi gas model} \\
Width fluctuation & \pbox{20cm}{widthmode 2: \\ Hofmann-Richert-Tepel-Weidenm\"{u}ller} & widthmode 1: Moldauer model   \\
Spin cut-off parameter & spincutmodel 2:$\sigma^2$= c$\sqrt{U/a}$ & spincutmodel 1: $\sigma^2$ = c a/\~{a} $\sqrt{U/a}$   \\
Vibrational enhancement & kvibmodel 1: $K_{vib}$=exp($0.00555A^{2/3}t^{4/3}$) & kvibmodel 2: $K_{vib}$=exp[$\delta$ s-($\delta$ U /t)]   \\
Spin distribution (PE) & \pbox{20cm}{preeqspin 2: the spin distribution \\from total level densities is adopted}     & preeqspin n  \\
Component exciton model & twocomponent: y & twocomponent: n  \\
Gamma-strength function     & strength 8: Gogny D1M HFB+QRPA~\cite{bib:24}  &  strength 2: Brink-Axel Lorentzian~\cite{bib:33} \\
\bottomrule
\end{tabular}
\end{table*}


From the table, it can be seen that, in the case of the pre-equilibrium model for example, the pre-equilibrium model 3 (denoted by the TALYS keyword $preeqmode$ \textbf{3}) which implies that the Exciton model (Numerical transition rates with optical model for collision probability) was used in our calculations in place of the default two-component Exciton model~\cite{bib:13}. Similarly, in the case of level density, the Back-shifted Fermi gas model ($ldmodel$ \textbf{2}) was preferred to the default model (Constant Temperature + Fermi gas model). While the level density model selected and the default TALYS level density model are all phenomenological level density models, they defer in that, in the case of the default level density model, the Constant Temperature Model (CTM) is used at low energies in combination with the Fermi gas model at high energies while in the case of the Back-shifted Fermi gas model, instead of using a constant temperature part, the model is expressed in terms of an effective excitation energy~\cite{bib:13,bib:20}. Also, even though some semi-microscopic optical (JLM) models were included in model variations, default phenomenological optical model potentials (OMP) as implemented in TALYS were selected. TALYS uses the local and global parameterisations of Koning and Delaroche~\cite{bib:23,bib:29a} as the default optical model. In the case of the compound nucleus calculations, the default TALYS model for the width fluctuation correction (WFC) is the Moldauer expression, however, in the case of this work, the Hofmann-Richert-Tepel-Weidenm\"{u}ller (HRTW) model was selected for the width fluctuation corrections (i.e TALYS keyword $widthmode$ \textbf{2}). It should be noted however that, using the HRTW model instead of Moldauer has no effect on proton induced reactions. For the strength function, the Gogny D1M HFB+QRPA~\cite{bib:24} ($strength$ \textbf{8}) was selected in place of the default Brink-Axel Lorentzian model~\cite{bib:33} ($strength$ \textbf{2}). 

\begin{figure*}[!htb] 
  \centering
  \includegraphics[trim = 20mm 70mm 20mm 70mm, clip, width=0.85\textwidth]{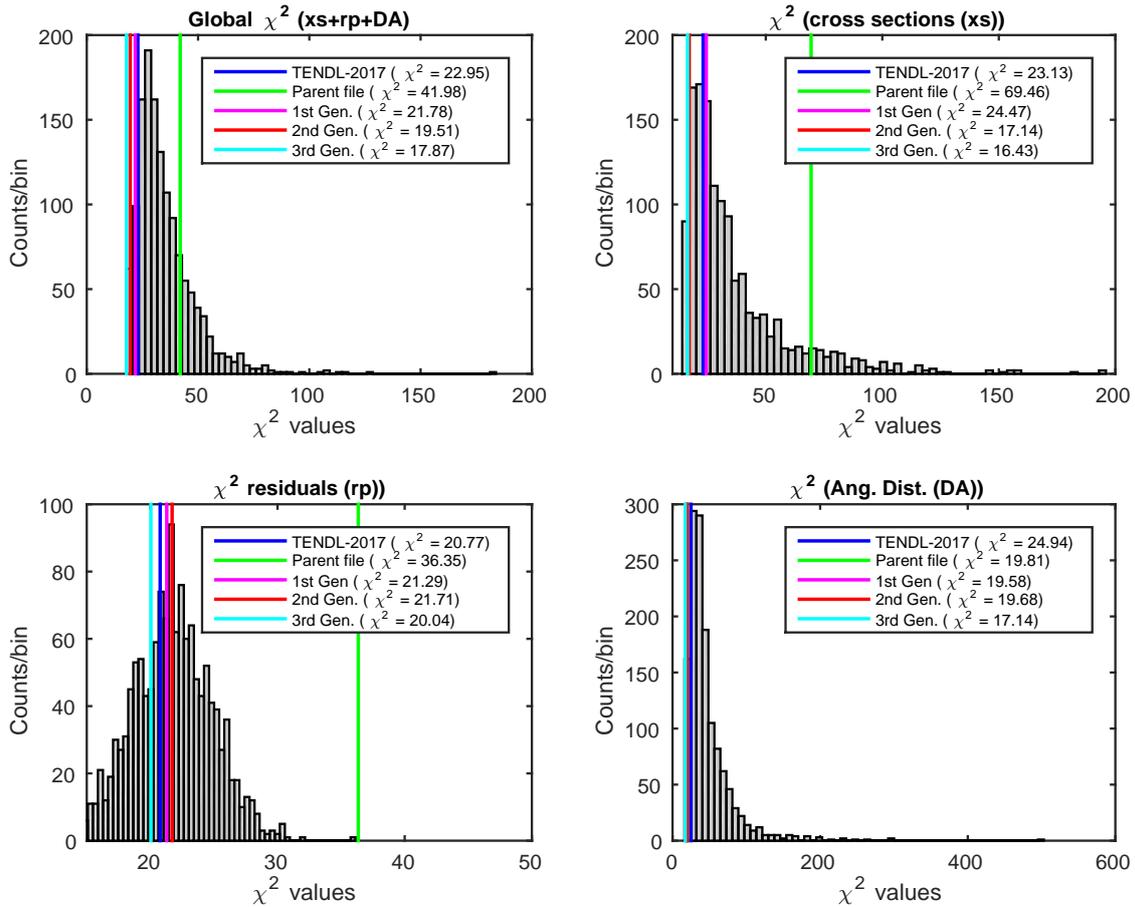} 
  \caption{Reduced $\chi^2$ distributions obtained from the 3rd generation are presented for the three experimental data types considered as well as the global averaged reduced $\chi^2$. The reduced chi squares computed for the parent, 1st, 2nd and 3rd generations are compared with TENDL-2017 library for each considered experimental type as well as the global reduced chi square. xs denotes reaction cross sections, rp – residual production cross sections and DA – angular distributions. The global reduced $\chi^2$ was computed by combining the individual reduced $\chi^2$ obtained from the different experimental data types.}
  \label{chi_sq_distr}
\end{figure*}

In Fig.~\ref{chi_sq_distr}, the global reduced $\chi^2$ distribution as well as the reduced $\chi^2$ distributions computed for the reaction cross sections, the residual production cross sections and the elastic angular distributions for p+$^{59}$Co, are presented. The $\chi^2$ distributions in the plots represent the distribution from the 3rd generation. Also in the same figure, the reduced chi square values computed for the 'best' file from each generation i.e. from the parent, 1st, 2nd and 3rd generations, are compared with the values obtained for the TENDL-2017 evaluation using the same experimental data. The global reduced $\chi^2$ distribution was obtained by taking the average over the different experimental types considered. 
%
%
%
Random variation of parameters were then made around the parent file to produce the 1st generation of random ND files from which the next 'best' file was obtained. This was repeated in an iterative fashion until a relative difference of 9.2\% (which is less than our 10\% target) was obtained between the global MLE estimates of the last two iterations. Normally, the iterations should be repeated several times until convergence however, it should be noted here that, the creation of random ND files (depending on the energy grid and computational resources available), can take several days or even weeks to complete and hence can be computationally expensive to consider several iterations. Furthermore, since we are normally interested in a target accuracy (for example, to improve on the previous TENDL evaluation as in this case), the iteration can be stopped if this objective is achieved.  

From Fig.~\ref{chi_sq_distr}, it can be seen that, our 'best' file from the parent generation did poorly compared with the other evaluations. This is expected since a smaller number of excitation energy bins of 20 was used for TALYS calculations in the case of the parent generation while a bin size of 60 was used for the other generations as well as for the TENDL evaluation. It was observed that, accuracy of TALYS calculations increases with the number of energy bins but at a higher computational cost. In the case of the global reduced  $\chi^2$, except for the parent file, the evaluations from the 1st, 2nd and 3rd generations outperformed the TENDL evaluation: reduced $\chi^2$ values of 41.98, 21.78, 19.51 and 17.87 were obtained for the parent, 1st, 2nd and 3rd generations compared with reduced $\chi^2$ values of 22.95 obtained for the TENDL evaluations. It can also be seen that, the results from the 1st generation is an improvement over the parent file as expected: average reduced $\chi^2$ values of 21.78, 24.47, 21.29, 19.58 were obtained for the global, reaction cross sections, residual production cross sections and elastic angular distributions respectively in the case of the 1st generation (1st Gen.) compared with 41.98, 69.46, 36.35, and 19.81 for the parent file respectively. Modest gains were however made with regards to the 2nd generation and 3rd generations: a global average $\chi^2$ value of 19.51 and 17.87 were obtained respectively, compared with 21.78 for the 1st generation. 

\begin{figure*}
  \centering
  \includegraphics[trim = 15mm 17mm 10mm 15mm, clip, width=0.40\textwidth]{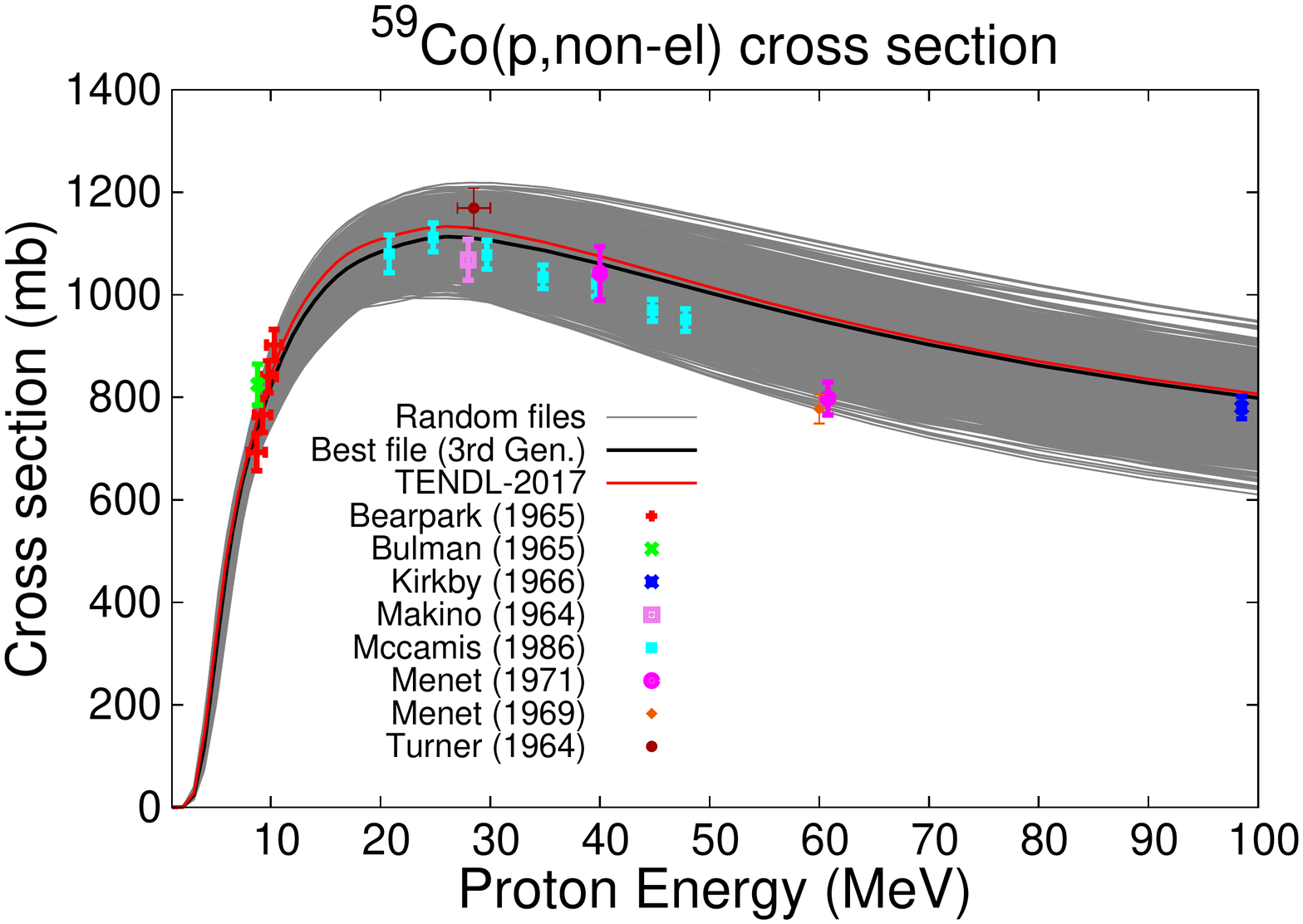}
  \includegraphics[trim = 15mm 17mm 10mm 15mm, clip, width=0.40\textwidth]{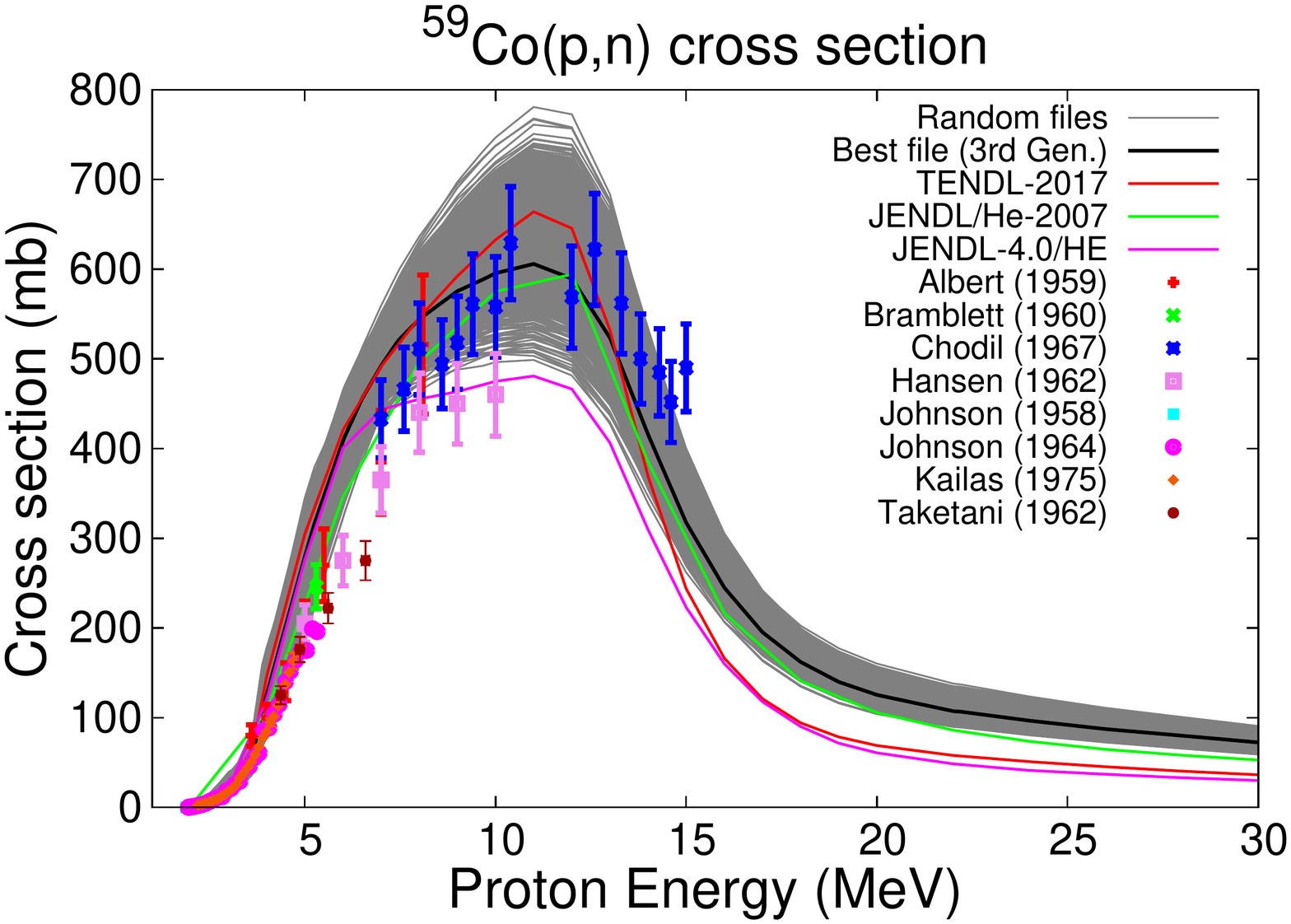}
  \includegraphics[trim = 15mm 17mm 10mm 15mm, clip, width=0.40\textwidth]{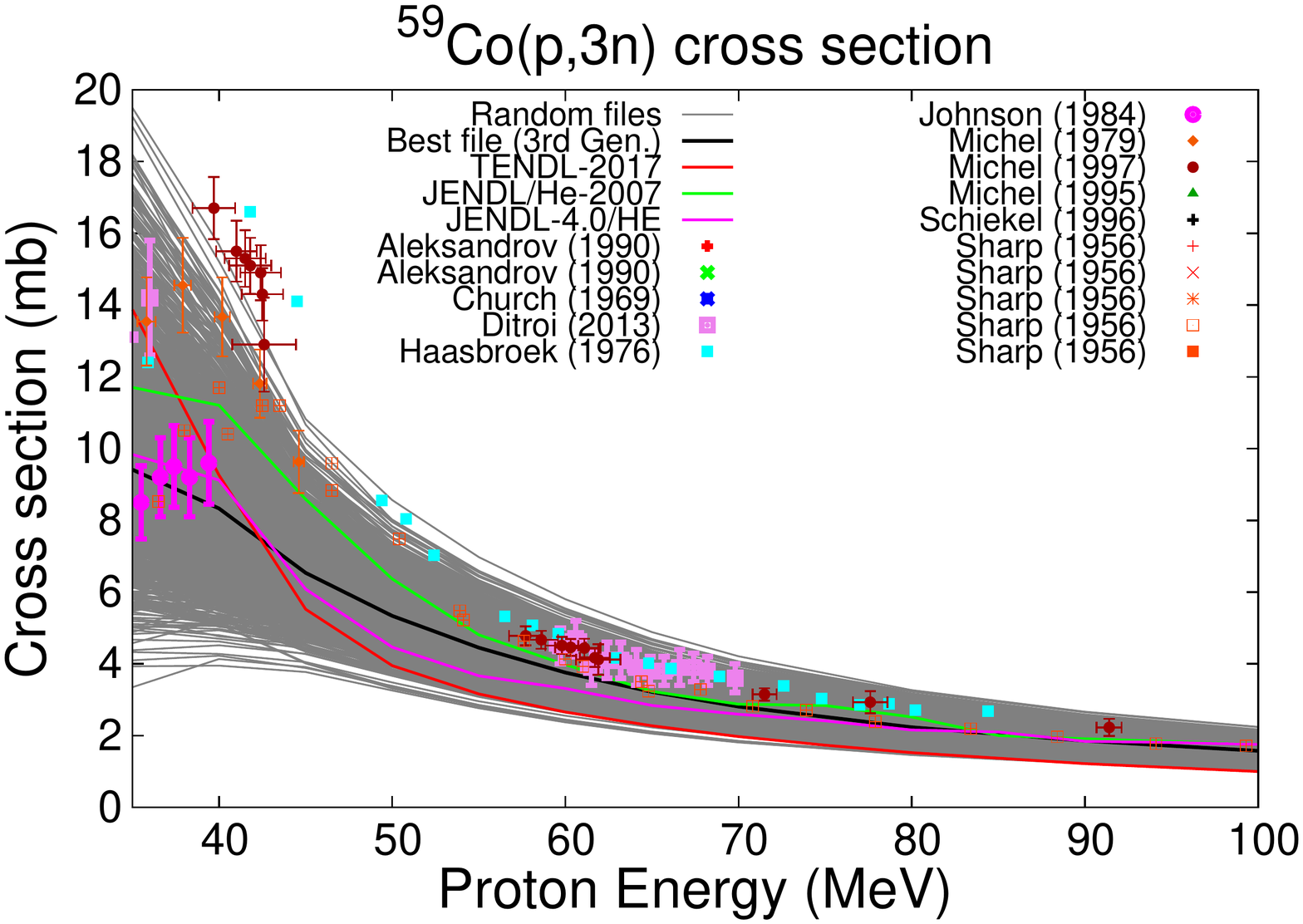}
  \includegraphics[trim = 15mm 17mm 10mm 15mm, clip, width=0.40\textwidth]{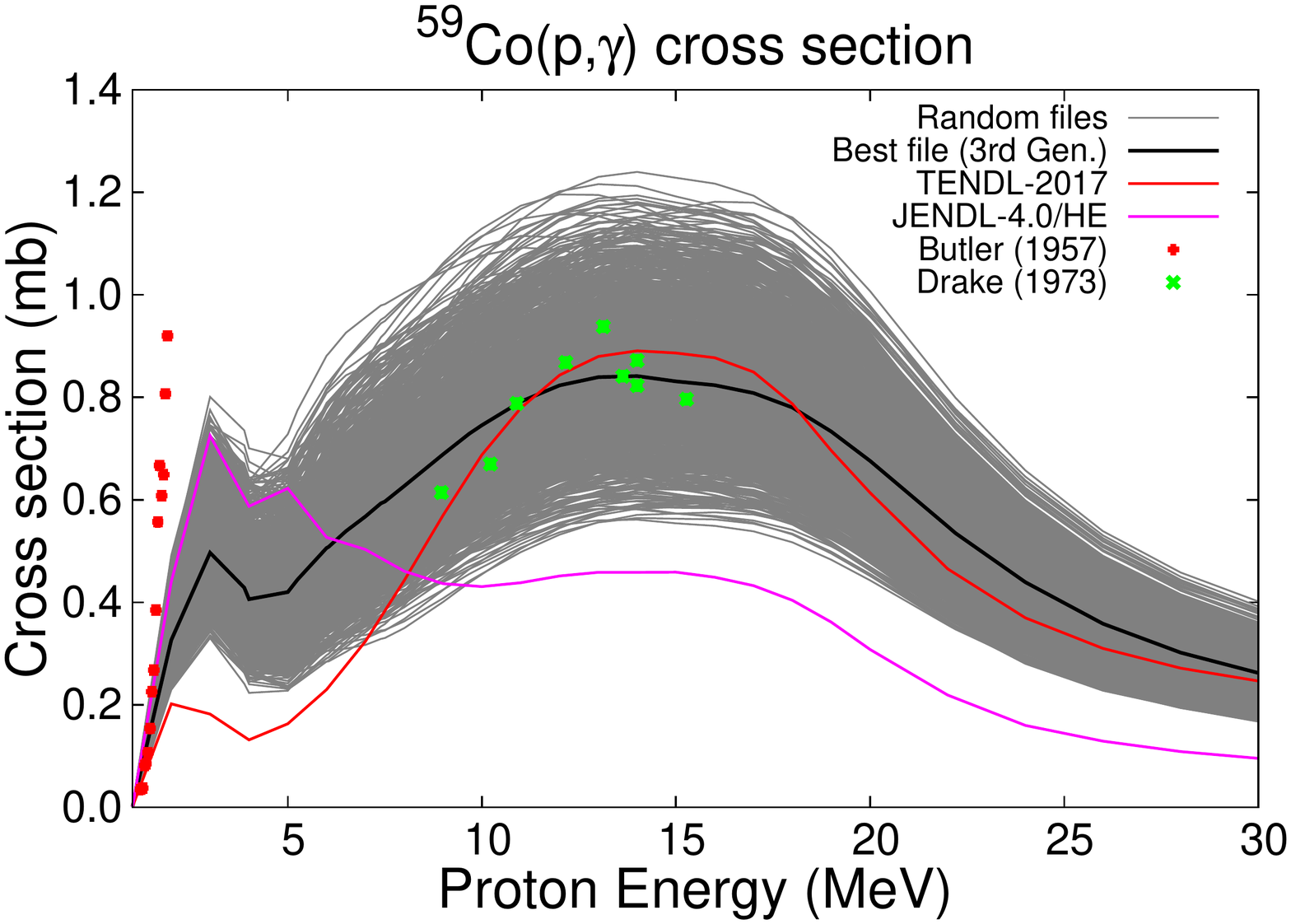} 
  \caption{Comparison between the evaluations from this work (i.e. Best file (3rd Gen.)) and the TENDL-2017 library for the (p,non-el), (p,n), (p,3n) and (p,$\gamma$) cross sections of p+$^{59}$Co. Also, comparisons are made with the JENDL/He-2007 and JENDL-4.0/HE libraries in cases where evaluations are available.}
  \label{file_performance_xs}
 
  
  \centering
  \includegraphics[trim = 15mm 17mm 10mm 15mm, clip, width=0.40\textwidth]{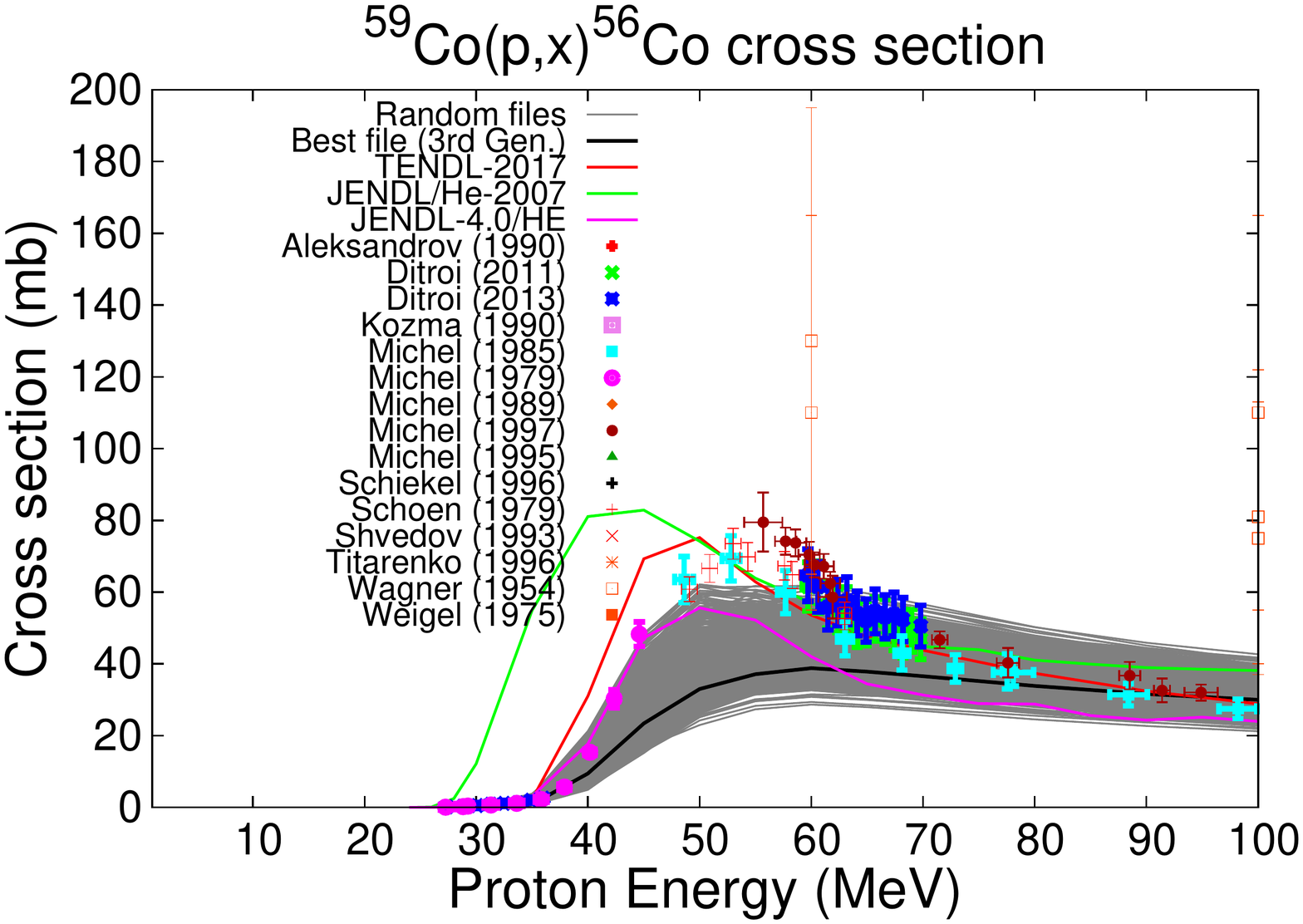}
  \includegraphics[trim = 15mm 17mm 10mm 15mm, clip, width=0.40\textwidth]{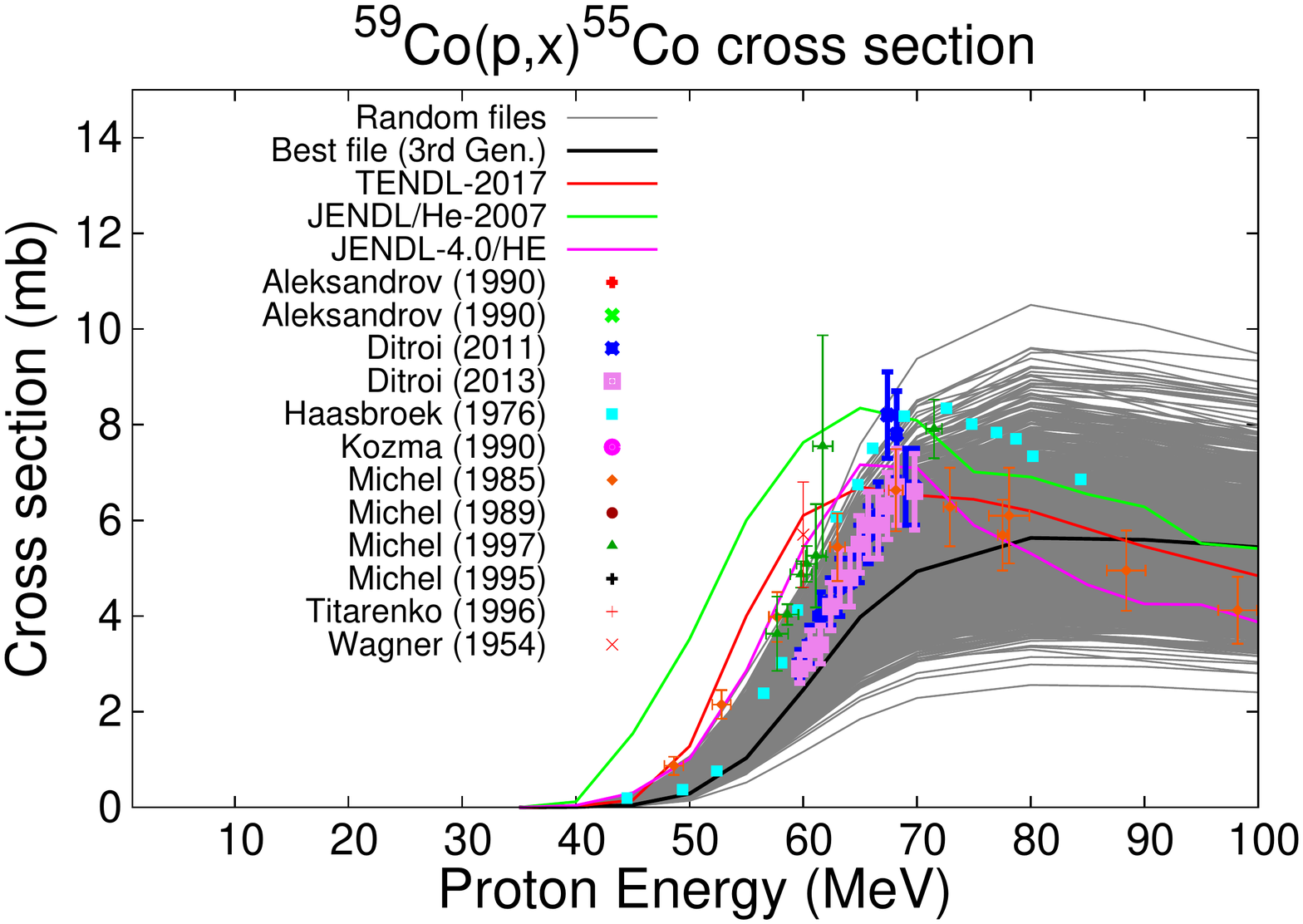}
  \includegraphics[trim = 15mm 17mm 10mm 15mm, clip, width=0.40\textwidth]{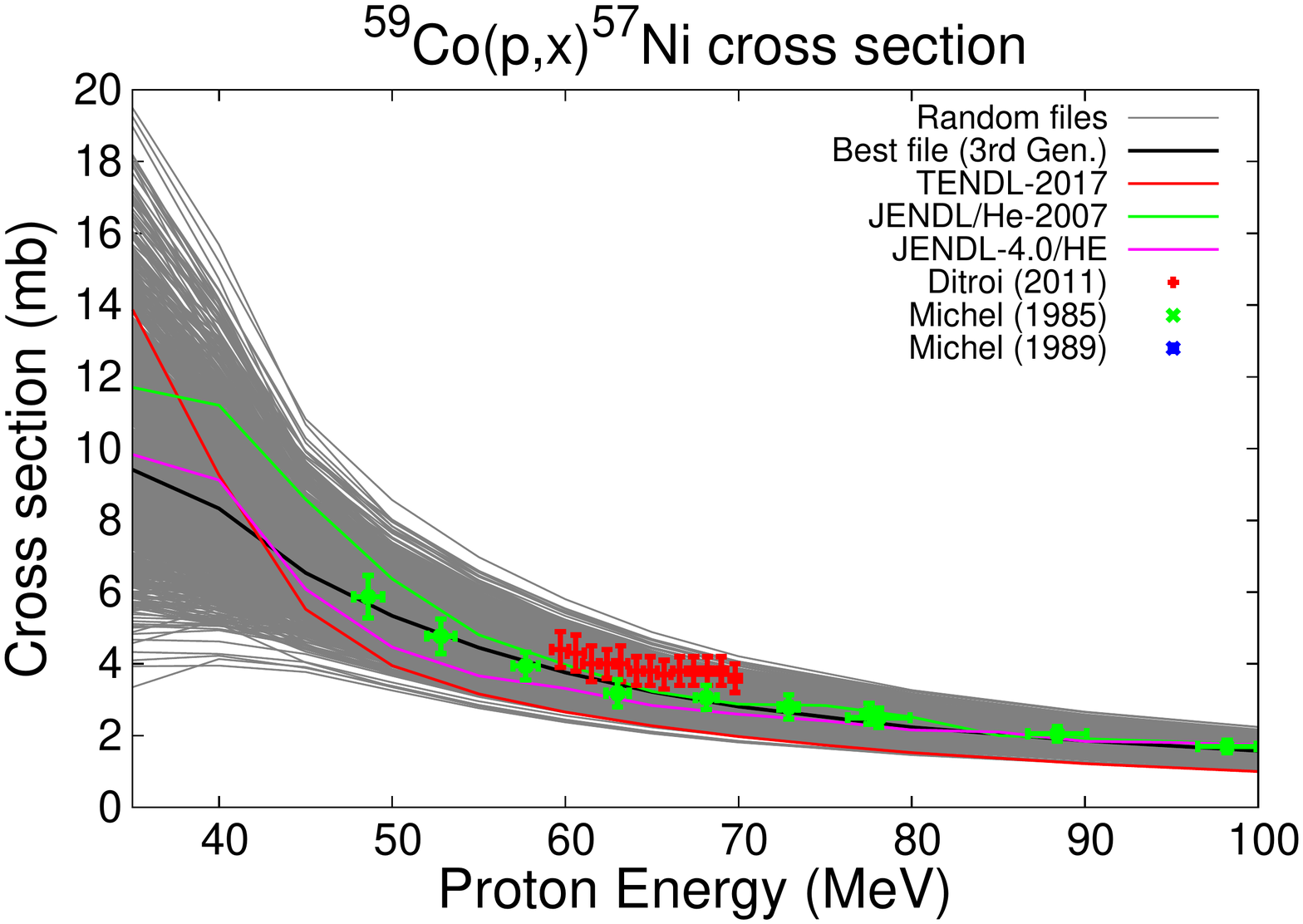}
  \includegraphics[trim = 15mm 17mm 10mm 15mm, clip, width=0.40\textwidth]{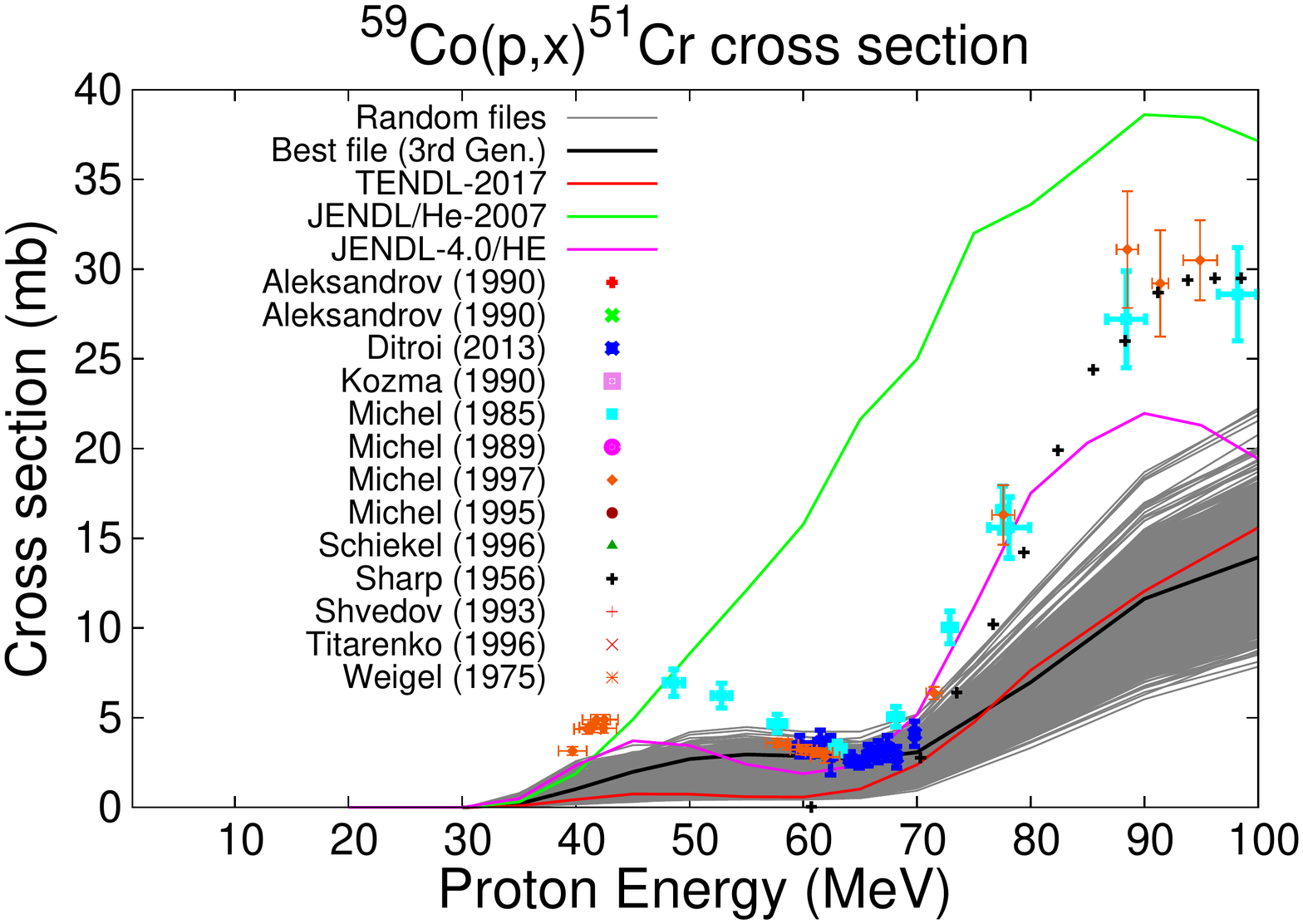} 
  \caption{Excitation functions against incident proton energies for the following residual cross sections: $^{59}$Co(p,x)$^{56}$Co, $^{59}$Co(p,x)$^{55}$Co, $^{59}$Co(p,x)$^{57}$Ni and $^{59}$Co(p,x)$^{51}$Cr. The evaluation from this work is compared with the evaluations from TENDL-2017, and the JENDL/He-2007 and JENDL-4.0/HE libraries.}
  \label{file_performance_rp}
  \end{figure*}

  In Tables~\ref{Exp_rxns_chi2_Co59} and \ref{Exp_rp_chi2_Co59}, the reduced chi squares values computed for the different channels used in the adjustments and computed for each generation are compared with evaluations from the TENDL-2017 and JENDL-4.0/HE ND libraries, in the case of the reaction and residual production cross sections respectively. Comparisons were made with the TENDL and the JENDL libraries because, only these libraries have evaluations for p+$^{59}$Co. The reduced chi square values given in the last column of Tables~\ref{Exp_rxns_chi2_Co59} and \ref{Exp_rp_chi2_Co59} were obtained by optimizing model calculations to only experimental reaction and residual production cross sections respectively. From Table~\ref{Exp_rxns_chi2_Co59}, it can be seen from the large reduced chi squares obtained for Gen. 0, 1, 2, and 3 that, it was difficult for the TALYS code to reproduce the (p,$\gamma$) cross section. It should however be noted that, it was impossible to cover the entire model and parameter space given the computational resources and time available and hence, there might exist better solutions that have not been identified in this work. In the case of the JENDL-4.0/HE library, a very large reduced chi square value of 1367.91 was obtained signifying that, the JENDL-4.0/HE evaluation was completely off compared to some experimental data. From Fig.~\ref{file_performance_xs} (bottom right of figure), it can be observed that, the JENDL-4.0/HE evaluation significantly under-predicts the the data from Drake (1973). A relatively smaller reduced chi square value of 40.33 was however obtained for the TENDL-2017 evaluation of the $^{59}$Co(p,$\gamma$) cross section. In the case of our evaluations, the relatively large reduced chi squares obtained for Gen. 0, 1, 2 and 3, is largely due to the inability of our models to reproduce some experimental data points from Butler (1957) as well as from Drake (1973). No experimental uncertainties were reported for these measurements (i.e. Drake (1973) and Butler (1957)), therefore, a 10\% relative uncertainty was assumed (for each data point) since these were the only experimental data sets available within the considered energy range. Also, no data were available for the (p,non), (p,np)g, and (p,np)m channels in the JENDL-4.0/HE library and therefore the average reduced chi square presented in Table~\ref{Exp_rxns_chi2_Co59} was averaged over the (p,n), (p,3n), (p,4n), (p,$\gamma$) and the (p,xn) reaction channels. 
  
  In Table~\ref{Exp_rp_chi2_Co59}, it can be seen that, the $^{59}$Co(p,x)$^{48}$V (rp023048 - TALYS name) benefited greatly from the iterative procedure; a large reduced chi square value of 172 for the parent generation was decreased to a value of 37.28 (Gen. 3). However, it can be observed that, the JENDL-4.0/HE evaluations outperformed the evaluations from this work and that from the TENDL-2017 library; reduced chi square values of 13.42 and 60.02 were obtained for the JENDL-4.0/HE and TENDL-2017 respectively. In fact the JENDL-4.0/HE library outperformed our evaluations except in the case of the $^{59}$Co(p,x)$^{57}$Ni, where our evaluation was in a better agreement with experimental data. As mentioned previously, one disadvantage of a multi-objective optimization procedure is that, it gives the best trade-off solutions, hence a situation can occur (as in this case), where even though our evaluation does perform better globally, it performed badly locally. In Table~\ref{Exp_DA_chi2_Co59}, a comparison of the reduced chi square values obtained for the different generations computed and that of the TENDL-2017 library for p+$^{59}$Co between 1-100 MeV are presented for the elastic angular distributions. No evaluation was available in the JENDL-4.0/HE and JENDL/He-2007 libraries for p+$^{59}$Co elastic angular distributions and therefore not presented. Also, with reference to the parent generation (Gen. 0), there were no TALYS results for the following incident energies: 5.25, 6.50, 7.00, 7.50, 30.30 and 40.00 MeV, and therefore, no results are reported for these energies. It should be noted however that, this did not have significant impact on the optimization since the elastic angular distributions did not vary much with model variations. Furthermore, it can be seen from the table that, the evaluations from the 3rd generation is a significant improvement on the parent generation. It can also be observed that, our evaluation outperformed the TENDL-2017 evaluation in this case. In addition, it can be observed from Tables~\ref{Exp_rxns_chi2_Co59}, \ref{Exp_rp_chi2_Co59} and \ref{Exp_DA_chi2_Co59} that, by using data from each experimental data type (one at a time) in our optimization, relatively smaller averaged reduced chi square values can be obtained. For example, improvements can be observed when these values are compared with the evaluations from the 3rd generation (Gen. 3). This is expected since in determining the optimal solution (in each case), we only took into consideration data from each experimental data type.  
  
  

  \begin{table*}
       \centering
       \caption{Comparison of the reduced chi square values between the different generations computed and the evaluations from the TENDL-2017 and JENDL-4.0/HE libraries for p+$^{59}$Co between 1-100 MeV, in the case of reaction cross sections. No cross section data were available for the (p,non), (p,np)g, and (p,np)m channels in the JENDL-4.0/HE library and therefore, not presented. The average value of 282.50 for the  JENDL-4.0/HE library, was obtained by taking the average over the (p,n), (p,3n), (p,4n), (p,$\gamma$) and (p,xn) channels. In the last column (only xs), the reduced chi squares were obtained by optimizing model calculations to only experimental reaction cross section data.}
       \label{Exp_rxns_chi2_Co59}
       \begin{tabular}{ccccccccc}  
       \toprule
       MT entry & \pbox{20cm}{Cross \\ section}  & Parent Gen. (Gen. 0)  & 1st Gen. (Gen. 1) & 2nd Gen. (Gen. 2) & 3rd Gen(Gen. 3) & TENDL-2017 & JENDL-4.0/HE & Only xs \\
       \midrule
       MT003 & (p,non) & 3.80 &  1.39 & 2.87 &  2.85 & 3.80 & - & 4.32 \\
       MT004 & (p,n) & 5.90 & 3.82 & 3.59 & 3.60 & 6.27 & 5.48 & 3.62 \\
       MT017 & (p,3n)  & 16.04 & 20.06 & 13.38 & 16.08 & 24.11 & 15.13 & 7.65 \\
       MT028g & (p,np)g & 2.22 & 0.85 & 1.06 & 1.62 & 4.77 &  - & 0.36\\
       MT028m & (p,np)m & 2.30 & 3.12 & 4.60 & 4.12 & 3.06 & - & 6.61 \\
       MT037 & (p,4n) & 9.63 & 7.57 & 4.36 & 5.36 & 14.45 & 21.36 & 2.28 \\
       MT102 & (p,$\gamma$) & 457.18 & 107.35 & 48.52 & 52.14 & 40.33 & 1367.91 & 49.53 \\
       MT201 & (p,xn) & 58.64 & 51.64 & 58.78 & 45.71 & 89.87 & 2.62 & 38.72 \\
       \hline
             & Average & 69.46 & 24.47 & 17.14 & 16.43 & 23.33 & 282.50 & 14.14 \\ 
    \bottomrule
    \end{tabular}

       \centering
       \caption{Comparison of the reduced chi square values between the different generations from this work and the evaluations from the TENDL-2017 and JENDL-4.0/HE libraries for p+$^{59}$Co between 1-100 MeV in the case of residual cross sections. See Table~\ref{Exp_data_residuals} for more information on the experimental data sets used in the computation of the reduced chi square presented. In the last column (only rp), the reduced chi squares were obtained by optimizing model calculations to only experimental residual production cross section data.}
       \label{Exp_rp_chi2_Co59}
       \begin{tabular}{ccccccccc}  
       \toprule
       TALYS name & \pbox{20cm}{Cross \\ section}  & Parent Gen. (Gen. 0)  & Gen. 1 & Gen. 2 & Gen. 3 & TENDL-2017 & JENDL-4.0/HE & Only rp \\
       \midrule
       rp021046 & $^{59}$Co(p,x)$^{46}$Sc  & 16.45 & 28.97 & 28.25 & 13.38 & 31.60 & 5.11 & 9.52  \\
       rp023048 & $^{59}$Co(p,x)$^{48}$V  &  172.00 & 49.28 & 39.81 & 37.93 & 60.02 & 13.42 & 17.81 \\
       rp025052 & $^{59}$Co(p,x)$^{52}$Mn  &  30.29 & 36.92 & 36.96 & 37.28 & 31.01 & 18.86 & 30.08 \\
       rp026055 &  $^{59}$Co(p,x)$^{55}$Fe &  24.57 & 17.47 & 22.15 & 15.47 & 15.08 & 7.10 & 13.22 \\
       rp027055 &  $^{59}$Co(p,x)$^{55}$Co &  2.99 & 12.07 & 13.85 & 17.13 & 7.24 & 4.10 & 3.88 \\
       rp027056 &  $^{59}$Co(p,x)$^{56}$Co  &  32.04 & 12.65 & 17.55 & 25.78 & 6.93 & 16.27 & 10.05 \\
       rp027057 &  $^{59}$Co(p,x)$^{57}$Co  & 12.17 & 11.80 & 14.78 & 13.08 & 2.91 & 1.85  & 13.05 \\
       rp028057 &  $^{59}$Co(p,x)$^{57}$Ni  & 0.27 & 1.16 & 0.34 & 0.26 & 9.21 & 1.14 & 0.79 \\
       \hline
             & Average & 36.35 & 21.29 & 21.71 & 20.04 & 20.50 & 8.48 & 12.30 \\ 
    \bottomrule
    \end{tabular}
    \end{table*}

   \begin{table*}
       \centering
       \caption{Comparison of the reduced chi square values between the different generations computed and the  TENDL-2017 evaluation for p+$^{59}$Co between 1-100 MeV in the case of the elastic angular distributions. Also, the angles considered were between 1 - 180 deg. In the last column (only DA), the reduced chi squares were obtained by optimizing model calculations to only experimental elastic angular distributions data. Note: In cases were there were no TALYS results for the considered incident energy, no results are reported.}
       \label{Exp_DA_chi2_Co59}
       \begin{tabular}{cccccccc}  
       \toprule
       \pbox{20cm}{Incident  \\ energy (MeV)} & \pbox{20cm}{Author of Exp.}  & Parent Gen. (Gen. 0)  & 1st Gen. (Gen. 1) & 2nd Gen. (Gen. 2) & 3rd Gen. (Gen. 3) & TENDL-2017 & only DA\\
       \midrule
       5.25 &  D.A. Bromley & - &  0.47 & 0.30 & 0.53 & 1.38 & 0.30 \\
       6.50 & K. Kimura & - & 4.63 & 4.27 & 3.93 & 5.60 & 4.42 \\
       7.00 & K. Kimura  & - & 6.54 & 6.45 & 5.92 & 9.97 & 6.77 \\
       7.40 & K. Kimura & 9.58 & 8.98 & 7.84 & 6.98 & 9.56 & 7.65 \\
       7.50 & W.F. Waldorf & - & 74.96 & 60.62 & 52.35 & 67.21 & 57.40 \\
       9.67 & G.W. Greenlees & 32.71 & 34.86 & 23.17 & 17.76 & 32.70 & 14.21 \\
       11.00 & C.M. Perey & 17.14 & 19.38 & 14.33 & 13.84 & 17.18 & 9.31 \\
       30.30 & B.W. Ridley & - & 20.27 & 49.82 & 41.85 & 56.60 & 35.10 \\
       40.00 & M.P. Fricke &  - & 6.14 & 10.32 & 11.09 & 24.94 & 6.72 \\
       \hline
             & Average & 19.81 & 19.58 & 19.68 & 17.14 &  22.95 & 15.76 \\ 
    \bottomrule
    \end{tabular}
    \end{table*}
    
  
 In Fig.~\ref{file_performance_xs}, excitation functions for the (p,non-el), (p,n), (p,3n) and (p,$\gamma$) cross sections of $^{59}$Co are presented and compared with the evaluations from the TENDL-2017 library. When available, comparisons are made also with evaluations from JENDL-4.0/HE as well as the older JENDL/He-2007 library. It can be observed from the figure that, our  evaluation (i.e. best file (3rd Gen.)) performed better than the TEND-2017 library for the (p,non-el), (p,n), (p,3n) cross sections. In the case of the (p,3n) channel for example, the TENDL-2017 evaluation under predicted the cross sections from about 45 to 100 MeV. The evaluation from this work, is observed to be in good agreement with experiments from Sharp (1956) and Ditroi (2013). From about 35 - 50 MeV, our evaluation appears to reproduce the measurements from Johnson (1984) even though this experimental data set was not used in the optimization. The measurements from Johnson (1984) were found not to be consistent with other measurements within the same energy range such as Michel (1997) and Michel (1979), and  therefore was not considered in the optimization. With respect to the (p,non-el) and (p,n) channels, our evaluation describes the experimental data reasonably well compared with the TENDL-2017 evaluation. The JENDL/He-2007 evaluation for the (p,n) channel however outperforms both the evaluations from this work and that from the TENDL-2017 and JENDL-4.0/HE libraries especially between about 5 - 12 MeV. The evaluation from JENDL-4.0/HE appears to be worse compared with the JENDL/He-2007 library which is an older library. One observation made in this work is that, it appears that more effort was put into improving the residual production cross sections in the JENDL-4.0/HE library compared with the reaction cross section.
 In general, with reference to the (p,$\gamma$) channel, our evaluation fitted poorly to some experiments from both Butler (1957) especially between 1.21 - 1.26 MeV and 1.6 - 1.9 MeV. 
 

 \begin{figure*}[t] 
  \centering
  \includegraphics[trim = 10mm 20mm 0mm 20mm, clip, width=0.9\textwidth]{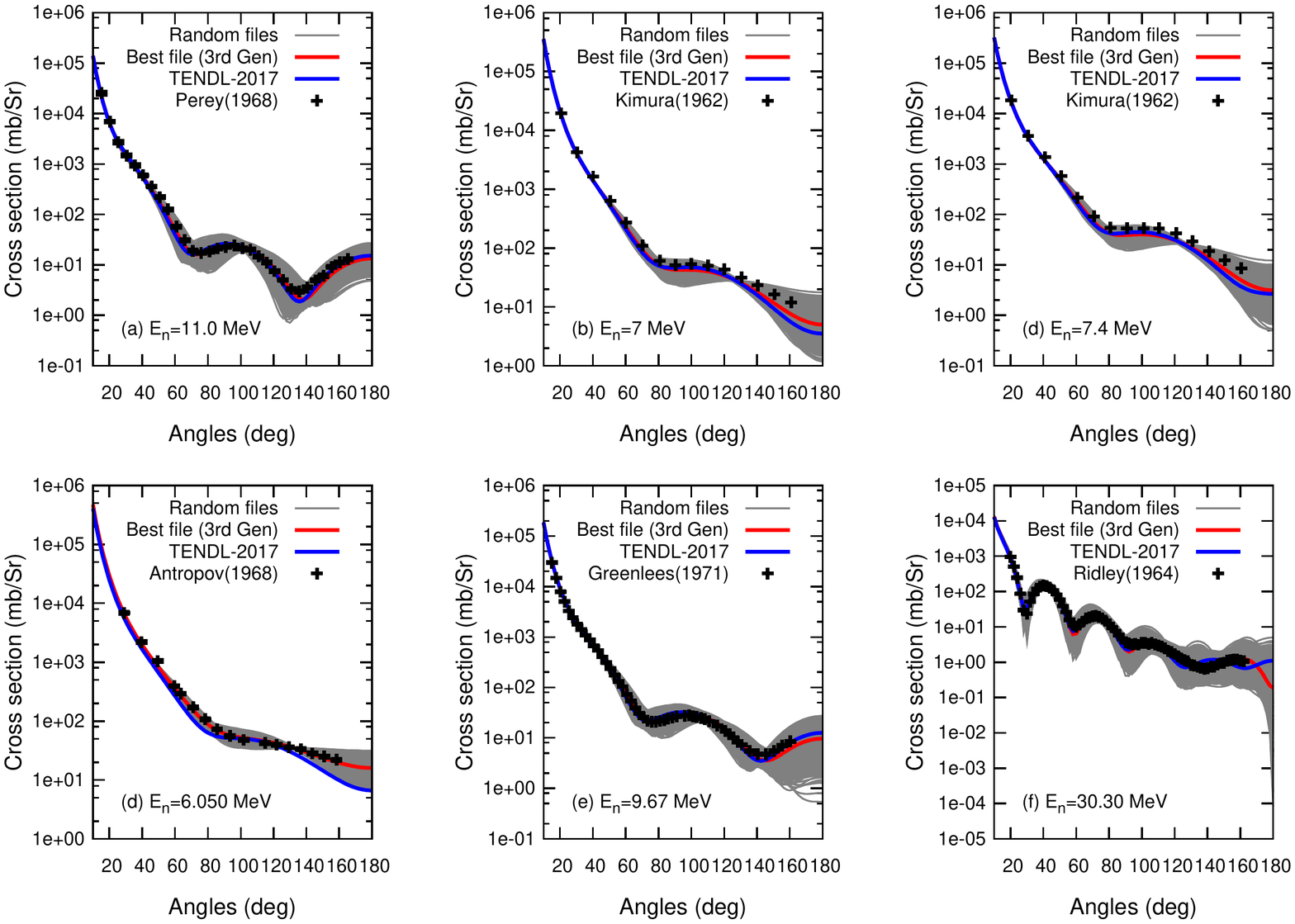} 
  \caption{Cross sections against angles (deg.) for selected incident energies: (a) 11.0 MeV, (b) 7 MeV, (c) 7.4 MeV, (d) 6.05 MeV, (e) 9.67 MeV and (f) 30.3 MeV for elastic angular distributions of p+$^{59}$Co. The evaluation from this work (Best file (3rd Gen.)) is compared with the TENDL-2017 evaluation. The TENDL evaluation was obtained by rerunning the TALYS code with the same model and parameter set used to create the TENDL-2017 evaluation. The plots in gray represent random cross sections from the 3rd generation.}
  \label{file_performance_da}
  \end{figure*} 
  
 In Fig.~\ref{file_performance_rp}, the cross sections against incident proton energies computed for the following residual cross sections: $^{59}$Co(p,x)$^{56}$Co, $^{59}$Co(p,x)$^{55}$Co, $^{59}$Co(p,x)$^{57}$Ni and $^{59}$Co(p,x)$^{51}$Cr are compared with the evaluations from the TENDL-2017, and the JENDL/He-2007 and JENDL-4.0/HE libraries. From the figure, it can be seen that our evaluation  (Best file (3rd Gen.)) and the evaluations from the JENDL-4.0/HE and JENDL/He-2007 libraries are in good agreement with experimental data with reference to the $^{59}$Co(p,x)$^{57}$Ni cross section. The TENDL evaluation under predicted this cross section for the entire energy range where experimental data are available. 
 %
 %
 The JENDL-4.0/HE library performed better for all the cross sections presented except in the case of the $^{59}$Co(p,x)$^{56}$Co where the TENDL-2017 evaluation performed better. This can be seen from the reduced chi squares presented in Table~\ref{file_performance_rp}: a reduced chi square of 16.27 was obtained for JENDL-4.0/HE while a reduced chi square of 6.93 and 25.78 were obtained for the TENDL-2017 evaluation and the (Gen. 3) respectively. All the libraries had difficulty reproducing experiments for the $^{59}$Co(p,x)$^{55}$Co between about 55 - 65 MeV. Our evaluation in the case of $^{59}$Co(p,x)$^{56}$Co under predicted the experimental data from about 40 MeV to 65 MeV but fits data from Michael (1985) from about 85 to 100 MeV. Similarly, in the case of the $^{59}$Co(p,x)$^{51}$Cr, our evaluation as well as the TEND-2017 and JENDL/He-2007 evaluations were unable to fit satisfactorily the experimental data available. In fact, the JENDL/He-2007 largely over predicted the cross section over the entire energy region. The JENDL-4.0/HE library which performed better, is a significant improvement on JENDL/He-2007 library in the case of the residual production cross sections. As seen from Fig.~\ref{file_performance_rp} (as well as from the reduced $\chi^2$ values for the residual production cross sections presented in Fig.~\ref{chi_sq_distr}), the TENDL-2017 evaluation globally out-performed our evaluation with respect to the residual production cross sections. This gives the indication that, the model set used for the TENDL-2017 library are better able to reproduce experimental data with respect to the residual production cross sections for p+$^{59}$Co. 
 
%
%
 %
In Fig.~\ref{file_performance_da}, the cross sections against angles (deg.) for selected incident energies: (a) 11.0 MeV, (b) 7 MeV, (c) 7.4 MeV, (d) 6.05 MeV, (e) 9.67 MeV and (f) 30.3 MeV for elastic angular distributions of p+$^{59}$Co from this work, are compared with the evaluation from the TENDL-2017 library. The TENDL evaluation was obtained by rerunning the TALYS code with the same model and parameter set used to create the TENDL-2017 evaluation but with an energy grid that contained incident energies of the experimental data. In this way, we were able to obtain perfect matches (in incident energy) between experiments and the TENDL-2017 and our evaluations. Globally, as seen from the reduced chi square values presented in Table~\ref{Exp_DA_chi2_Co59}, our evaluation (Best file (3rd Gen.)) outperformed the TENDL-2017 evaluation. From Table~\ref{Exp_DA_chi2_Co59}, an average reduced chi square of 17.14 was obtained for our evaluation while a value of 22.95 was obtained for the TENDL-2017 library. Over the entire angle range, our evaluation fits satisfactorily to the experimental data except in the high angles region where some deviations were observed. For example, it can be observed from Fig.~\ref{file_performance_da} that, our evaluation as well as TENDL-2017, under predicted the cross sections from about 140 - 180 degrees for the incident energies $E_n$ = 6.050, 7 and 7.4 MeV. 

 
  %

\subsection{Application to p+$^{111}$Cd}
In Table~\ref{models_selected_Cd111}, the 'best' models selected for the p+$^{111}$Cd case, are presented. These models were selected by comparing the likelihood functions computed for each model combination using the experimental cross sections presented in Table~\ref{Exp_data_Cd111}. 

 \begin{table*}[t]
 \centering
  \caption{The 'best' model combination selected in the case of p+$^{111}$Cd. These models were used as the new 'best' file around which model parameters were varied to obtain the 1st generation outputs. Other models not presented here were set to default values in the 'best' file. For the vibrational enhancement of the level density ($K_{vib}$) as presented, $\delta S$ and $\delta U$ are the changes in the entropy ($S$) and excitation energy ($U$), respectively, $t$ is the thermodynamic temperature and $A$ denotes the mass number~\cite{bib:14,bib:18,bib:19}. In the case of the spin cut-off parameter ($\sigma^2$), $U$ is the excitation energy, $a$ is energy-dependent level density parameter, $\~{a}$ is the asymptotic level density value obtained when all shell effects are damped and $c$ is the rigid body moment of inertia~\cite{bib:14,bib:18}. Gogny D1M HFB represents the Hartree-Fock-Bogolyubov model with the Gogny D1M nucleon force while QRPA is the Quasi-particle Random Phase Approximation model~\cite{bib:24}. $y$ and $n$ represents yes and no}.
  \label{models_selected_Cd111}
  \begin{tabular}{lll}  
  \toprule
   Model name/type  & Selected models & default models \\
   \midrule
Pre-equilibrium & \pbox{20cm}{preeqmode 3: Exciton model - Numerical \\ transition rates with optical model \\ for collision probability}   &  \pbox{20cm}{preeqmode 2: Exciton model: Numerical \\ transition rates with energy-dependent \\ matrix element}  \\
Level density & ldmodel 2: Back-shifted Fermi gas model &  \pbox{20cm}{ldmodel 1: Constant temperature \\ + Fermi gas model} \\
Width fluctuation & widthmode 1: Moldauer model  & widthmode 1: Moldauer model   \\
Spin cut-off parameter & spincutmodel 1: $\sigma^2$ = c a/\~{a} $\sqrt{U/a}$  & spincutmodel 1: $\sigma^2$ = c a/\~{a} $\sqrt{U/a}$   \\
Vibrational enhancement & kvibmodel 2: $K_{vib}$=exp[$\delta$ s-($\delta$ U /t)] & kvibmodel 2: $K_{vib}$=exp[$\delta$ s-($\delta$ U /t)]   \\
Spin distribution (PE) & \pbox{20cm}{preeqspin 3: the spin distribution \\ is based on particle-hole state densities}  & preeqspin n  \\
Component exciton model & twocomponent: y & twocomponent: y  \\
Gamma-strength function & strength 5: Goriely’s hybrid model~\cite{bib:25} & strength 2: Brink-Axel Lorentzian \\
 Surface corrections (PE)   & preeqsurface n  & preeqsurface y \\
 Liquid drop expression & shellmodel 2: Expression by Goriely~\cite{bib:33}  & shellmodel 1: Expression by Myers-Siatecki~\cite{bib:33} \\
 Spherical Optical Model & spherical n  & spherical y \\
\bottomrule
\end{tabular}
\end{table*}

\begin{table*}
       \centering
       \caption{Comparison of the reduced chi square values between the different generations from this work and the evaluation from the TENDL-2017 for p+$^{111}$Cd between 1-100 MeV. The following nuclear reaction channels were considered: (p,n), (p,2n), (p,2n)m, (p,2n)g, (p,3n) and (p,4n). See Table~\ref{Exp_data_Cd111} for more information on the experimental data used in the computation of the chi squares.}
       \label{Exp_rxns_chi2_Cd111}
       \begin{tabular}{cccccc}  
       \toprule
       MT entry & \pbox{20cm}{Cross \\ section}  & Parent Gen. (Gen 0)  & 1st Gen. (Gen. 1)  & 2nd Gen. (Gen. 2) & TENDL-2017 \\
       \midrule
       MT004 & (p,n) & 37.25 & 9.86 & 6.63 & 22.43  \\
       MT016 & (p,2n)  & 3.25 & 2.83 & 3.57 & 3.54  \\
       MT016m & (p,2n)m  & 15.94 & 16.12 & 12.74 & 2.30  \\
       MT016g & (p,2n)g  & 3.69 & 2.74 & 2.25 & 1.64  \\
       MT017 & (p,3n) & 20.12 & 14.41 & 8.73 & 16.57 \\
       MT037 & (p,4n) & 20.27 & 11.41 & 4.64 & 13.48  \\
       \hline
             & Average & 16.75 & 9.56 & 6.43 & 10.02 \\ 
    \bottomrule
    \end{tabular}
    \end{table*}

\begin{figure*}[t] 
  \centering
  \includegraphics[trim = 15mm 17mm 10mm 15mm, clip, width=0.40\textwidth]{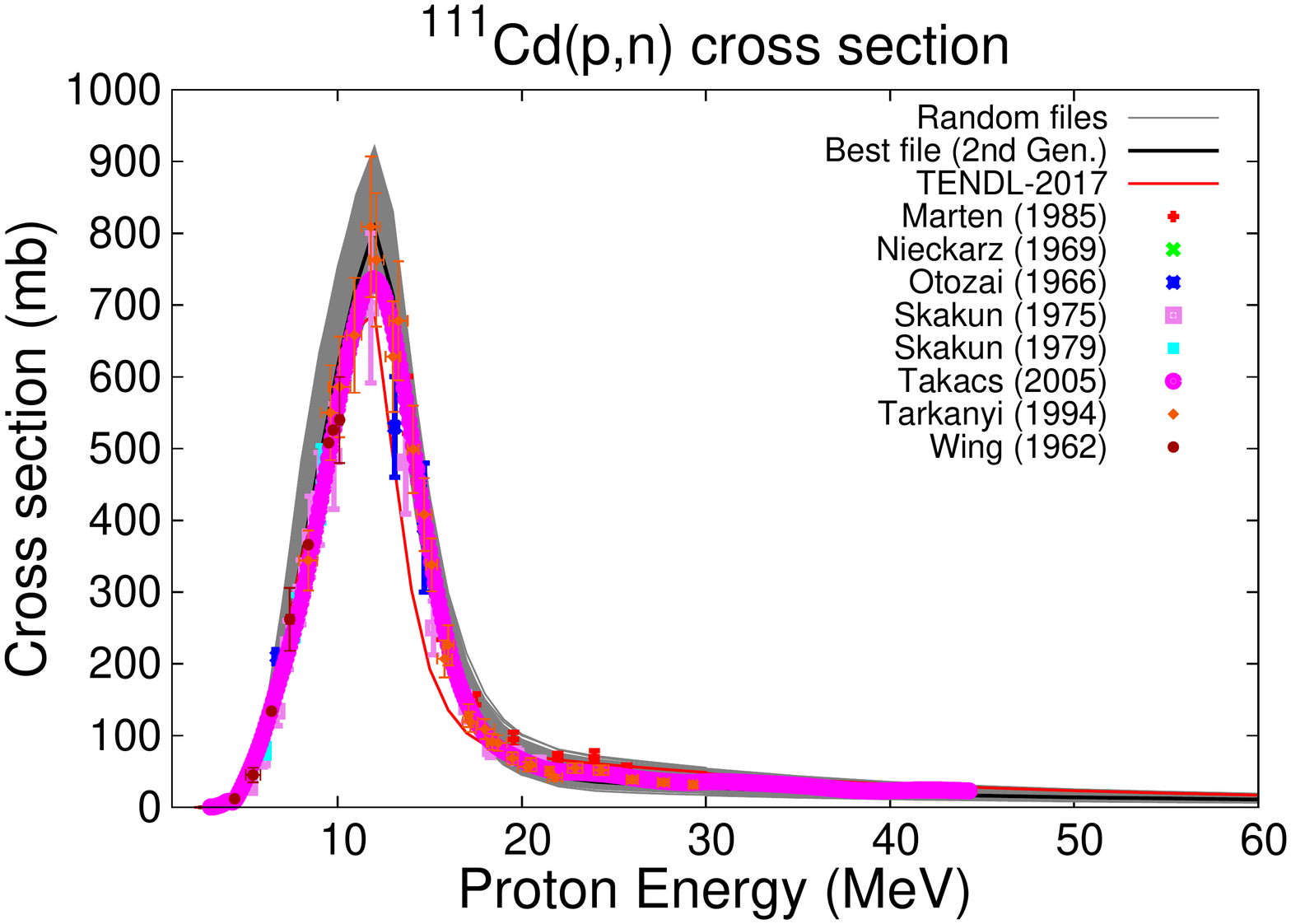}
  \includegraphics[trim = 15mm 17mm 10mm 15mm, clip, width=0.40\textwidth]{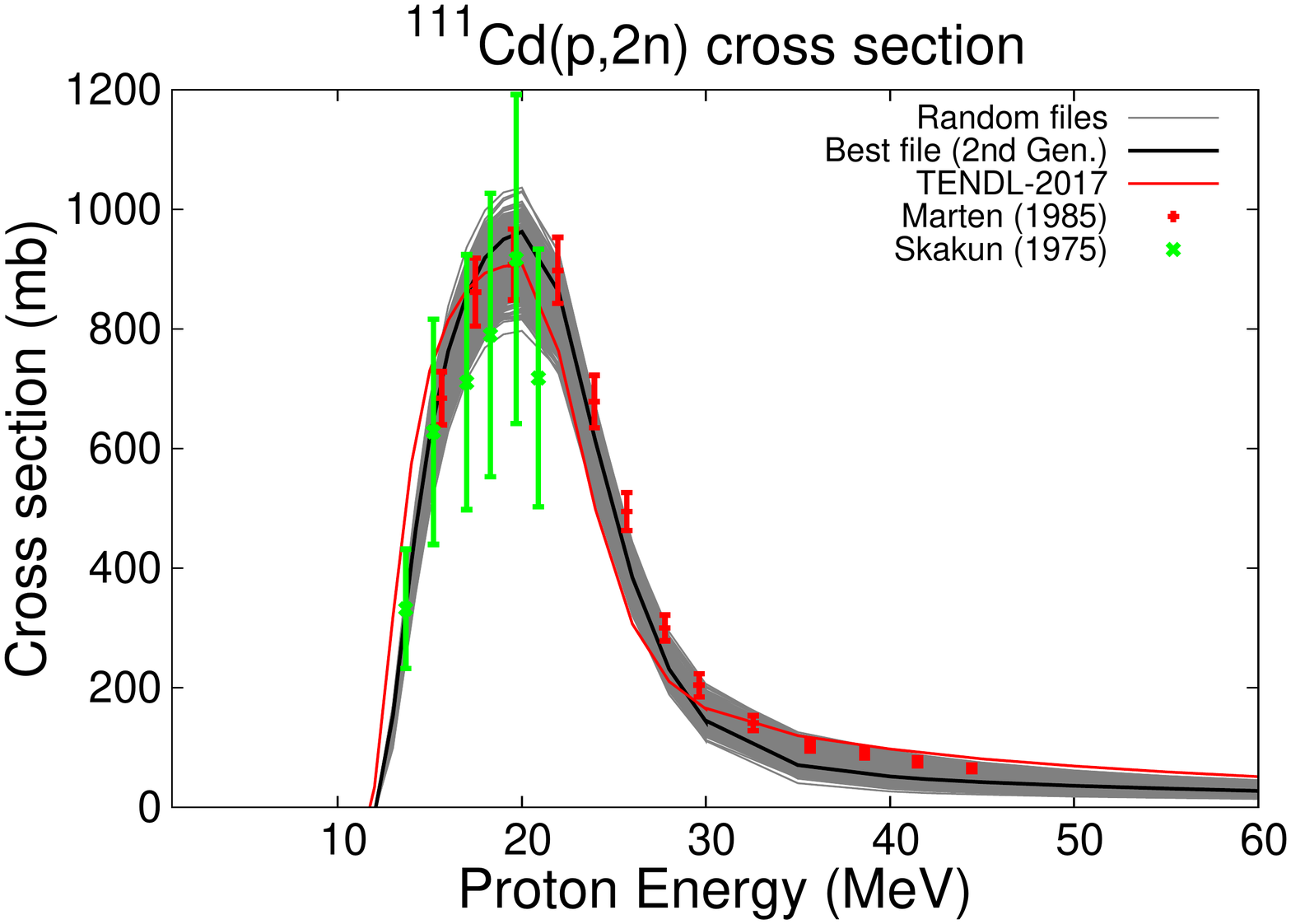}
  \includegraphics[trim = 15mm 17mm 10mm 15mm, clip, width=0.40\textwidth]{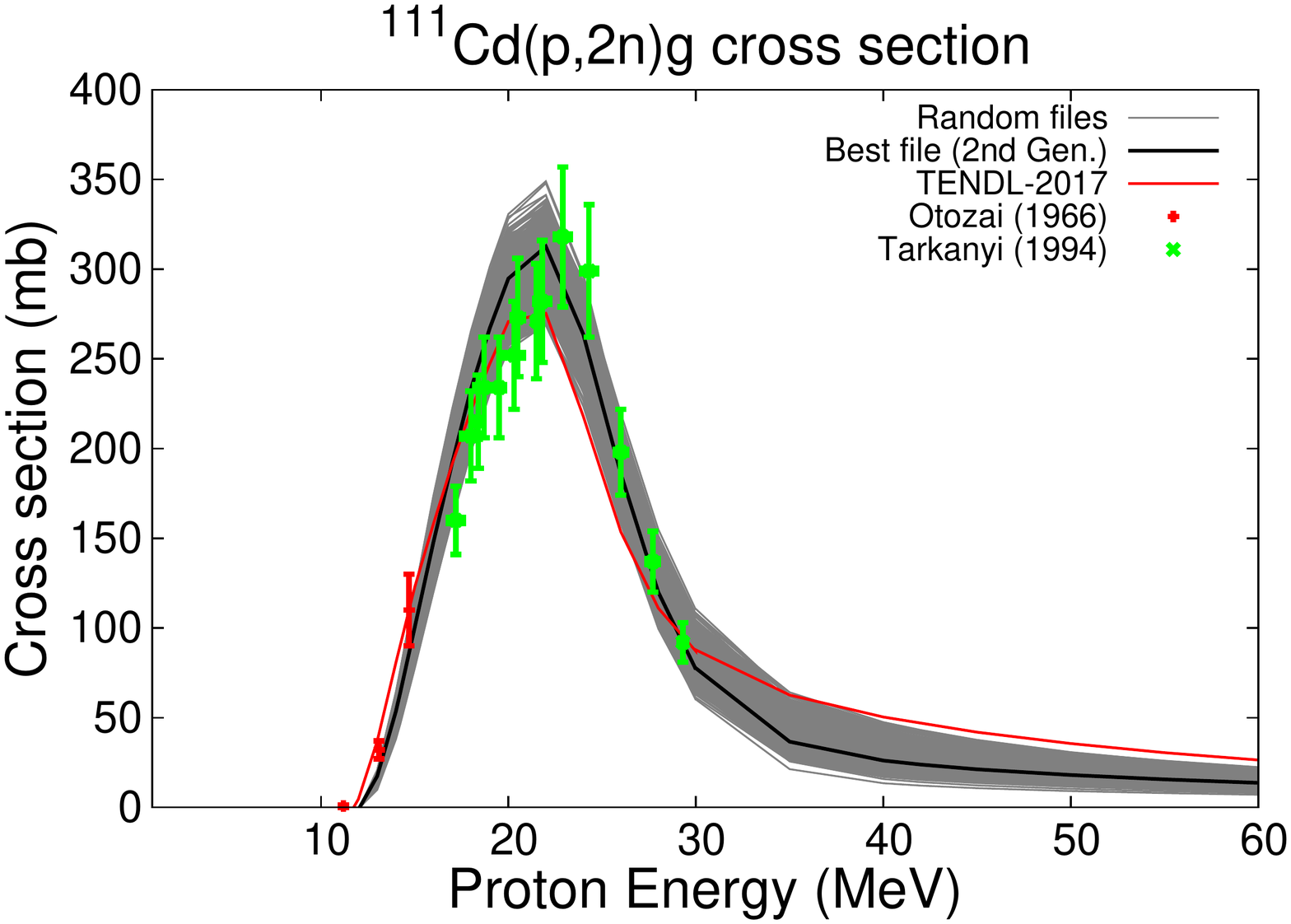}
  \includegraphics[trim = 15mm 17mm 10mm 15mm, clip, width=0.40\textwidth]{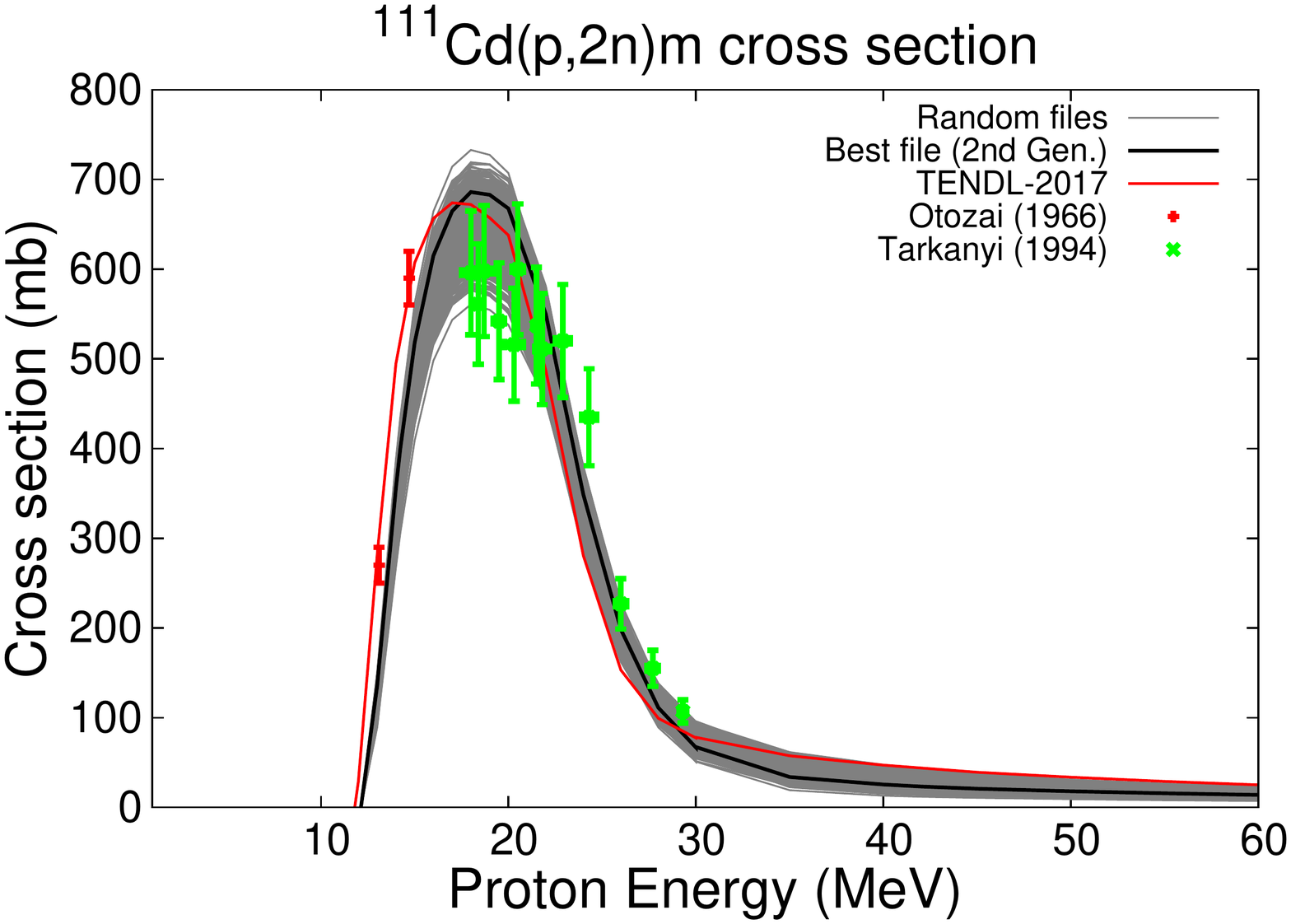} 
  \caption{Comparison of file performance between the evaluations from this work and the TENDL-2017 library for the (p,n), (p,2n), (p,2n)g and (p,2n)m cross sections of $^{111}$Cd. The plots in gray represent random cross sections from the 2nd generation. Note that, there were no evaluations for p+$^{111}$Cd in the ENDF/B-VIII.0, JENDL/He-2007 and the JENDL-4.0/HE libraries.}
  \label{xs1_Cd111}
\end{figure*}

From the table, it can be seen that, for example, $spherical$ \textbf{n} was used in our selected model combination while $spherical$ \textbf{y} is used for default TALYS calculations. When this flag is activated, it ensures that a spherical Optical Model Potential (OMP) calculation is carried out regardless of the availability of a deformed OMP and a coupling scheme~\cite{bib:14}. In the case of the pre-equilibrium model, similar to the p+$^{59}$Co case, the Exciton model (Numerical transition rates with optical model for collision probability) was selected instead of the default two-component Exciton model. In the case of the level density models, out of the six options available in the TALYS code, the Back-shifted Fermi gas model was selected. The default level density model implemented in TALYS is $ldmodel$ \textbf{1}: Constant temperature + Fermi gas model. Out of the 8 Gamma-strength function implemented in TALYS, Goriely’s hybrid model (TALYS keyword: $strength$ \textbf{5}) as presented in Ref.~\cite{bib:25} was selected in place of the Brink-Axel Lorentzian (TALYS keyword: $strength$ \textbf{2})~\cite{bib:33}. In nuclear reaction calculations, the gamma-ray transmission coefficients are key components used in the description of the gamma emission channel. This is particularly important because, in some reactions, gamma rays are emitted together with other particles~\cite{bib:25}. The parameters to the models (as presented in Table~\ref{models_selected_Cd111}), were varied all-together to produce a new set of random nuclear data files referred to as the 1st generation in this work. 

In Table~\ref{Exp_rxns_chi2_Cd111}, the reduced chi square values computed using experimental reaction cross section data (presented in Table~\ref{Exp_data_Cd111}), are presented for different generations and compared with the TENDL-2017 evaluation between 1 - 100 MeV. The p+$^{111}$Cd reaction is a classical case which shows the importance of  model variation. For example, it was difficult to reproduce the experimental data for the $^{111}$Cd(p,n) channel by varying only model parameters. However, a good description of experimental data was obtained for the (p,n) cross section after varying several models together (with their parameters with a number of iterations (see Fig.~\ref{xs1_Cd111} and Table~\ref{Exp_rxns_chi2_Cd111})). The $^{111}$Cd(p,n) reaction is important because, it is used for the production of $^{111}$In, a $\gamma$-emitter used in diagnostic nuclear medicine~\cite{bib:32}. From Table~\ref{Exp_rxns_chi2_Cd111}, it can be observed that, improvements in cross sections were achieved with each iteration except in the case of the (p,2n)m and the (p,2n) channels where the reduced chi square computed for the parent generation was found to be better than that of the 2nd generation. In the case of the (p,2n) channel, the parent and 1st generations performed better than the 2nd generation. Also, from the average reduced chi square, it is seen that our evaluation (2nd Gen.) globally performed better than the TENDL-2017 evaluation. However, when we look at the individual channels, the TENDL-2017 evaluation described the experimental data from Otozai (1966) better with respect to the (n,2n)m and (n,2n)g channels. This explains why smaller reduced chi squares were obtained for TENDL-2017 library compared to our evaluation (see Table~\ref{Exp_rxns_chi2_Cd111}). Since in the computation of the reduced chi square, we averaged over each experimental data set, the experiments from Otozai (1966) and Tarkanyi (1994) carried equal weights (aside their uncertainties) regardless of the number of measurements for each data set.

In Fig.~\ref{xs1_Cd111}, a comparison of file  performance between the evaluations from this work and the evaluation from the TENDL-2017 library are presented for the following cross sections of $^{111}$Cd: (p,n), (p,2n), (p,2n)g and (p,2n)m. In the case of the (p,n) channel, our evaluation outperformed the TENDL-2017 evaluation for the entire energy range. It can be seen also that, our evaluation describes the experiments reasonably well. 
For the (p,2n) channel, our evaluation is in good agreement with the experiments from Marten (1985) from about 15 to 30 MeV while the TENDL evaluation fits better, the experiments from Skakun (1975) which has rather relatively large experimental uncertainties. Good agreements were observed between our evaluation and the experimental data from both Tarkanyi (1994) and Otozai (1966) in the case of the  (p,2n)g channel. With reference to the (p,2n)m channel, our evaluation fitted better to the experimental data from Tarkanyi (1994) especially in the high energy region (between about 24 - 30 MeV). The TENDL evaluation however, is in good agreement with the experiments from Otozai (1966) as well as from Tarkanyi (1994). This explains why a relatively smaller reduced chi square of 2.30 was obtained for the TENDL-2017 evaluation of the (p,2n)m channel compared with 12.74 obtained for this evaluation (Gen. 2). In the case of (p,3n) and (p,4n) cross sections (not shown), it was observed that, our evaluations and that of TENDL-2017 under predicted the (p,3n) cross section between about 30 - 50 MeV,  and (p,4n), between about 40 and 50 MeV.  
%

%



  

\section{Conclusion}
In this work, we explored the use of an Iterative Bayesian Monte Carlo procedure for the improvement of nuclear data evaluations in the fast energy region. The goal of the iterative procedure was to minimize the difference between our experimental observables and the corresponding model outputs in an iterative fashion. This was done by exploring both the model and parameter space in order to identify the 'best' model and parameter sets that make our experimental data most probable within a Bayesian Monte Carlo framework. The associated uncertainties of the selected file can be obtained by re-sampling model parameters around this file. In this work, because we were unable to cover the entire model and parameter space, the default parameter uncertainties and distribution used for the parent generation was maintained for subsequent generations. For future work, the parameter uncertainties (and their distributions) should be updated for each iteration/generation. The method was applied for the evaluation of proton induced reactions on $^{111}$Cd and $^{59}$Co between the 1 - 100 MeV energy region. The study shows that, there is a potential for the improvement of nuclear data evaluations (within the limit of the available models), through an iterative process. It should be noted however that, multi-objective optimizations (as in our case) normally give rise to a set of trade-off solutions called Pareto-optimal solutions where each objective (with respect to each experimental data type as used), cannot be improved without degrading the quality of one or all of the other objectives. As a rule of the thumb, when no further improvements are possible, the iteration should be stopped since the Pareto optimum might have been reached. Alternatively, subjective weights could be assigned to each experimental data type depending on the needs of the evaluations. In this way, the optimization algorithm would favour the experimental data types with the larger weights. Furthermore, because of the computational cost involved, the entire model (and parameter) space could not covered in this work. It is therefore recommended that, more efficient sampling methods be explored in future work. Also, since the selected models were still observed to be deficient in their ability to reproduce some experimental cross sections, it is recommended that the effect of model defects be included in the iteration process in future work.


\section{Acknowledgment}
This work was done with funding from the Paul Scherrer Institute through the NES/GFA-ABE Cross Project. The authors would like to thank Georg Schnabel from the IAEA and H. Sj\"ostrand from Uppsala University for interesting discussions on this topic.


\begin{appendices}
\setcounter{table}{0}
\renewcommand{\thetable}{A\arabic{table}}

\section{Appendix}
In the Appendix, we present the experimental data from the EXFOR database used in the optimization procedure. In Tables~\ref{Exp_data_angle}, \ref{Exp_data_rxns} and \ref{Exp_data_residuals}, the selected experimental data for the elastic angular distributions, the reaction cross sections and the residual production cross sections with reference to p+$^{59}$Co, are presented respectively. In Table~\ref{Exp_data_Cd111}, the experimental data for the reaction cross section in the case of p+$^{111}$Cd are presented.

\begin{table}[h!]
       \centering
       \caption{Selected microscopic data for $^{59}$Co elastic angular distributions used for adjustments, showing the number of data points, the EXFOR ID and the name of the first author, for given incident energies. Note, only angles between 1 and 180 degrees were considered.}
       \label{Exp_data_angle}
       \begin{tabular}{ccccc}  
       \toprule
       Energy (MeV) &  \pbox{20cm}{Total number of  \\ data points} & Author  & EXFOR ID  & Year \\
       \midrule
       5.25 & 17 & D. A. Bromley & C1086002 & 1956 \\
       6.50 & 11 & K. Kimura  & E2232017 & 1962 \\
       7.00 & 7 & K. Kimura  & E2232018 & 1962 \\
       7.40 & 9 & K. Kimura  & E2232019 & 1962 \\
       7.50 & 19 & W. F. Waldorf & C1088003 & 1957 \\
       9.670 & 24 & G. W. Greenlees & O03930041 & 1971 \\
       11.00 & 18 & C. M. Perey & C2165007 & 1968 \\
       30.30 & 41 & B. W. Ridley & O0142006 & 1964 \\
       40.00 & 39 & M. P. Fricke & O0328004 & 1967 \\
       \bottomrule
       \end{tabular}
 \end{table}
%
%
 %
%

  \begin{table*}[hb]
       \centering
       \caption{Selected microscopic experimental data used for adjustments of p+$^{59}$Co in the case of the reaction cross sections showing the number of data points, the EXFOR ID and the name of the first author of the measurements. Note that, even though the energy range of each measurement are presented, only data points between 1 and 100 MeV were considered in this evaluation.}
       \label{Exp_data_rxns}
       \begin{tabular}{ccccccc}  
       \toprule
       MT entry & \pbox{20cm}{Cross \\ section}  &  \pbox{20cm}{Total number of data \\ points considered}  & Energy range   & Author  & EXFOR ID  & Year  \\
       \midrule
       MT003 & (p,non) &  4  &  8.72 - 10.3 MeV  & K. Bearpark & D0314004 & 1965  \\
       MT003 & (p,non) &  1  & 98.5 Mev  & P. Kirkby  & O0340012 & 1966  \\
       MT003 & (p,non)  &  1 &  28 MeV & M. Q. Makino &  C1212010 & 1964  \\
       MT003 & (p,non) & 7 &  20.8 - 47.8 MeV  & R. H. Mccamis & T0100006 & 1986  \\
       MT004 & (p, inl) & 25 & 1.89 - 2.46 MeV  & C. H. Johnson  &  T0122007 &  1958  \\
       MT004 & (p, inl) & 7  &  3.60 - 8.10 MeV  & R. D. Albert & T0130006 & 1959  \\
       MT004 & (p, inl)  & 1 &  5.30 MeV   & R. L. Bramblett  &  F1108019 & 1960  \\ 
       MT004 & (p, inl)  & 15 &  7.00 - 15.0 MeV  & G. Chodil & C06930052 & 1967 \\
       MT017 & (p,3n)  & 18 & 16.0 - 69.8 MeV  & F. Ditroi  & D4293002 & 2013 \\
       MT017 & (p,3n)  & 7 & 25.4 - 44.6 MeV  & R. Michael  & A0146018 & 1979 \\
       MT017 & (p,3n)  & 17 & 39.7 - 278 MeV  & R. Michael  & O0276095 & 1997 \\
       MT017 & (p,3n)  & 23 & 27.5 - 99.3 MeV & R. A. Sharp  & P0034005 & 1956 \\
       MT028g & (p,np)g & 12 & 13.4 - 31.0 MeV & R. A. Sharp & P0034023 & 1956 \\
       MT028m & (p,np)m & 12 & 13.4 - 31.0 MeV  &  R. A. Sharp & P0034022 & 1956 \\
       MT037 & (p,4n) &   10 & 59.7 - 69.8 MeV & F. Ditroi & D4293003 & 2013 \\
       MT037 & (p,4n) &   10 & 39.7 - 91.4 MeV  & R. Michael & O0276094 & 1997 \\
       MT102 & (p,$\gamma$) &  15  &  1.00 - 1.90 MeV   & J. W. Butler   &  P0036002  &  1957 \\
       MT102 & (p,$\gamma$) &   8  & 7.02 - 22.1 MeV  & D. M. Drake &  O0092003 & 1973 \\
       MT201 & (p,xn) & 16 &  7.00 - 15.0 MeV & G. Chodil & C06930051 & 1967 \\
    \bottomrule
    \end{tabular}
 \end{table*}

 \begin{table*}[!htb]
       \centering
       \caption{Selected microscopic experimental data for the production cross sections of residual nuclei showing the number of data points, the EXFOR ID and information on the authors of the measurements. Note that only the first authors have been listed. Also, even though the energy range of each measurement are presented, only data points between 1 and 100 MeV were considered in this work.}
       \label{Exp_data_residuals}
       \begin{tabular}{ccccccc}  
       \toprule
       TALYS name & \pbox{20cm}{Cross \\ section}  &  \pbox{20cm}{Total number of  \\ data points used}  & Energy range   & Author  & EXFOR ID  & Year \\
       \midrule
       rp021046 & $^{59}$Co(p,x)$^{46}$Sc  &  5  &  72.9 - 199 MeV & R. Michael  & A01000105 & 1985 \\
       rp023048 & $^{59}$Co(p,x)$^{48}$V  &  4 & 71.5 - 154 MeV   & R. Michael  & O0276086  & 1997 \\
       rp023048 &  $^{59}$Co(p,x)$^{48}$V & 13 & 63.8 - 98.8 MeV  & R. A. Sharp     & P0034021 & 1956 \\
       rp025052 & $^{59}$Co(p,x)$^{52}$Mn  & 13 & 59.7 - 69.8 MeV   & F. Ditroi & D4255008 & 2011 \\
       rp026055 &  $^{59}$Co(p,x)$^{55}$Fe & 25 & 21.8 - 84.4 MeV   & F. J. Haasbroek  & B00980031 & 1976 \\
       rp027055 & $^{59}$Co(p,x)$^{55}$Fe   & 10 &  48.6 - 166 MeV & R. Michael  & A01000095  & 1985 \\
       rp027055 & $^{59}$Co(p,x)$^{55}$Fe & 7  & 57.7 - 154 MeV & R. Michael  & O0276090  & 1997 \\
       rp027055 &  $^{59}$Co(p,x)$^{55}$Fe & 1  & 60.0 - 240 MeV & G. D. Wagner   & C0278004  & 1954 \\
       rp027055 &  $^{59}$Co(p,x)$^{55}$Co & 10 &  48.6 - 166.0 MeV  & R. Michel & A01000095 & 1985 \\
       rp027055 &  $^{59}$Co(p,x)$^{55}$Co & 7 & 57.7 - 154.0 MeV & R. Michael & O0276090 & 1997 \\
       rp027056 &  $^{59}$Co(p,x)$^{56}$Co  & 13 & 59.7 - 69.8 MeV & F. Ditroi & D4255004 & 2011 \\
       rp027056 & $^{59}$Co(p,x)$^{56}$Co   & 20 & 27.4 - 69.8 MeV  & F. Ditroi & D4293006 & 2013 \\
       rp027056 &  $^{59}$Co(p,x)$^{56}$Co  & 10 & 48.6 - 199. MeV  & R. Michael  & A01000094 & 1985 \\
       rp027056 &  $^{59}$Co(p,x)$^{56}$Co  & 13 & 30.8 - 154 MeV  & R. Michael  & O0276091 & 1997 \\
       rp027056 &  $^{59}$Co(p,x)$^{56}$Co  & 7 & 33.2 - 63.0 MeV & N. C. Schoen & T0276012 & 1979 \\
       rp027057 &  $^{59}$Co(p,x)$^{57}$Co  & 13 & 59.7 - 69.8 MeV   & R. Ditroi & D4255003 & 2011 \\
       rp027057 &  $^{59}$Co(p,x)$^{57}$Co  & 18 &  15.0 - 69.8 MeV   & R. Ditroi & D4293005 & 2013  \\
       rp027057 &  $^{59}$Co(p,x)$^{57}$Co  & 7  &  14.9 - 44.6 MeV   & R. Michel & A0146021 & 1979 \\
       rp027057 &  $^{59}$Co(p,x)$^{57}$Co  & 20 & 12.6 - 154.0 MeV   & R. Michel & O0276092  & 1997 \\
       rp027058 &  $^{59}$Co(p,x)$^{58}$Co  & 11 & 11.7 - 25.9 MeV   & A. A. Alharbi & D0673002 & 2011 \\
       rp027058 &  $^{59}$Co(p,x)$^{58}$Co  & 13 & 59.7 - 69.8 MeV & F. Ditroi  & D4255002 & 2011 \\
       rp027058 &  $^{59}$Co(p,x)$^{58}$Co  & 16 & 13.8 - 29.5 MeV  & V. N. Levkovski & A0510059  & 1991 \\
       rp027058 &  $^{59}$Co(p,x)$^{58}$Co  & 8 & 48.6 - 199.0 MeV & R. Michel & A01000092 & 1985  \\
       rp028057 &  $^{59}$Co(p,x)$^{57}$Ni  & 10 & 48.6 - 199.0 MeV   & R. Michel & A01000091 & 1985 \\
       \bottomrule
       \end{tabular}
       \end{table*}

  \begin{table*}
       \centering
       \caption{Selected microscopic experimental data used for the  adjustment of p+$^{111}$Cd in the case of the reaction cross sections showing the number of data points, the EXFOR ID and the name of the first author of the measurements. Note that, even though the energy range of each measurement are presented, only data points between 1 and 100 MeV were considered in the evaluation.}
       \label{Exp_data_Cd111}
       \begin{tabular}{ccccccc}  
       \toprule
       MT entry & \pbox{20cm}{Cross \\ section}  &  \pbox{20cm}{Total number of data \\  points considered}  & Energy range   & Author  & EXFOR ID  & Year  \\
       \midrule
       MT004 & (p, n) & 1 & 70 MeV  & W. J. Nieckarz  &  C0345002 &  1969  \\
       MT004 & (p, n) & 413  &  2.10 - 44.3  MeV  & S.Takacs & D41470041 & 2005  \\
       MT016 & (p, 2n)  & 13 &  13.8 - 44.44 MeV   & M. Marten  &  A0335004 & 1985  \\ 
       MT016 & (p, 2n)  & 6 &  11.8 - 20.9  MeV  & E. A. Skakun & A0001005 & 1975 \\
       MT016g & (p, 2n)g  & 2 &   11.2 - 14.7 MeV  & K. Otozai & P0019008 & 1966 \\
       MT016g & (p, 2n)g  & 14 &  13.3 - 29.3 MeV  & F. Tarkanyi & D4027003 & 1994 \\
       MT016m & (p, 2n)m  & 2 &  13.1 - 14.7 MeV  & K. Otozai & P0019007 & 1966 \\
       MT016m & (p, 2n)m  & 13 &  13.3 - 29.3 MeV  & F. Tarkanyi & D4027003 & 1994 \\
       MT017 & (p,3n)  & 10 &  22.0 - 44.4 MeV  & M. Marten  & A0335003 & 1985 \\
       MT037 & (p,4n) &   4 & 35.7 - 44.4 MeV  & M. Marten & A0335002 & 1985 \\
    \bottomrule
    \end{tabular}
    \end{table*}

 \end{appendices}

\end{document}